\documentstyle[preprint,eqsecnum,aps]{revtex}
\tightenlines
\input epsf
\newcommand{\piccie}[2]{\epsfxsize=#1 \epsfbox[10 30 560 590]{./#2}}

\begin{document}
\draft
%\preprint{HEP/123-qed}

\title{Implications of maximal Jarlskog invariant and maximal CP violation}

\author{E. Rodr\'{\i}guez-J\'auregui}
\address{Deutsches Elektronen Synchroton DESY, Hamburg \\ and \\
  Instituto de F\'{\i}sica, UNAM, Apdo. Postal 20-364, 01000 M\'exico,
  D.F. M\'exico.}

\date{\today}
\maketitle
\begin{abstract}    
We argue here why CP violating phase $\Phi$ in the quark mixing matrix is maximal, that is, $\Phi=90^{\circ}$. In the Standard Model CP violation is related to the Jarlskog invariant $J$, which can be obtained from non commuting Hermitian mass matrices. In this article we derive the conditions to have Hermitian mass matrices which give maximal Jarlskog invariant $J$ and maximal CP violating phase $\Phi$. We find that all squared moduli of the quark mixing elements have a singular point when the CP violation phase $\Phi$ takes the value $\Phi=90^{\circ}$. This special feature of the Jarlskog invariant $J$ and the quark mixing matrix is a clear and precise indication that CP violating Phase $\Phi$ is maximal in order to let nature treat democratically all of the quark mixing matrix moduli. 
\end{abstract}

\narrowtext
\section{Introduction}\label{sec:I}
In the Standard Model there is a single source of CP violation, that is a CP violating phase $\Phi$, in consequence of which all the CP observables are strongly correlated. The CP violation phase is introduced as a free parameter in the standard model. From the point of view of the fundamental theory of quarks and leptons this situation is not satisfactory and we would like to see the theoretical origin of the quark mixing matrix phase $\Phi$ which describes CP violation.\\
The purpose of the present article is to show that maximal Jarlskog invariant $J$ and maximal CP violating phase $\Phi$ are phenomenologically viable. The Jarlskog invariant $J$ is related to CP violation and is a function of the CP violating phase $\Phi$. It is well known that the Jarlskog invariant $J$ is obtained from the quark mass matrices ${\bf M}_{q}$. Without loss of generality we can consider quark mass matrices ${\bf M}_{q}$ to be Hermitian. The Jarlskog invariant for Hermitian mass matrices ${\bf M}_{q}$ is given as \cite{ref:4}, \cite{ref:35}
\begin{equation}\label{eq:I.3}
J=Im (det[{\bf M}_{d}, {\bf M}_{u}])/2F
\end{equation}
where
\begin{equation}\label{eq:I.5}
F=\left( 1+ \tilde{m}_{2u} \right)\left(1-\tilde{m}_{1u}\right)\left(\tilde{m}_{2u}+\tilde{m}_{1u}\right)\left(1+\tilde{m}_{2d}\right)\left(1-\tilde{m}_{1d}\right)\left(\tilde{m}_{2d}+\tilde{m}_{1d}\right),
\end{equation}
here $\tilde{m}_{iq}$ are the quark mass ratios $\tilde{m}_{iq}=m_{iq}/m_{3q}$ with subscripts $i=1, 2$ referring to $u, c$ in the u-type sector and $d, s$ in the d-type sector, $m_{3u}$ is the top quark mass and $m_{3d}$ is the beauty  quark mass.
For three quark families, ${\bf M}_{q}$ are $3 \times 3$ Hermitian mass matrices,

\begin{equation}\label{eq:1.7}
{\bf M}_{q}=m_{3q} \pmatrix{
E_q & A_q e^{-i\phi^{q}_a} & F_qe^{-i\phi^{q}_f} \cr
A_q e^{i\phi^{q}_a} & D_q & B_qe^{-i\phi^{q}_b} \cr
F_qe^{i\phi^{q}_f} & B_qe^{i\phi^{q}_b} & C_q \cr
}\qquad\qquad q=~u,~d.
\end{equation}
 The Hermitian mass matrix ${\bf M}_{q}$ may be written in terms of a real symmetric matrix ${\bf\bar M}_{q}$ and a diagonal matrix of phases ${\bf P}_{q}$ as follows

\begin{equation}\label{eq:1.9}
{\bf M}_{q}={\bf P}_{q}{{\bf\bar M}_{q}}{{\bf P}_{q}}^{\dagger},
\end{equation}
where
\begin{equation}\label{eq:1.11}
{\bf\bar M}_{q}=m_{3q}\pmatrix{
E_q & A_q  & F_q \cr
A_q  & D_q & B_q \cr
F_q & B_q & C_q \cr
}\quad\quad q=~u,~d,
\end{equation}
and
\begin{equation}\label{eq:1.13}
{\bf P}_{q}=diag(e^{i\phi^q_1},~e^{i\phi^q_2},~e^{i\phi^q_3}).
\end{equation}
 The phases in the Hermitian matrix ${\bf M}_{q}$ may be written as,
\begin{eqnarray}\label{eq:1.14}
\phi^{q}_a=(\phi^{q}_2-\phi^{q}_1),\qquad
\phi^{q}_f=(\phi^{q}_3-\phi^{q}_1),\qquad
\phi^{q}_b=(\phi^{q}_3-\phi^{q}_2),
\end{eqnarray}
that is, $\phi^{q}_a,~\phi^{q}_b$ and $\phi^{q}_f$ are correlated.
 The real symmetric matrix ${\bf\bar M}_{q}$ may be brought to a diagonal form by means of an orthogonal transformation 
\begin{equation}\label{eq:1.15}
{\bf\bar M}_{q}={\bf O}_{q}{\bf M}_{q, diag}{\bf O}^{T}_{q},\quad\quad q=~u,~d,
\end{equation}
where  ${\bf O}_{q}$ are orthogonal matrices and
\begin{equation}\label{eq:1.17}
{\bf M}_{q, diag}=m_{3q}~diag\left[ ~\tilde m_{1q},~\tilde m_{2q}, ~1\right].
\end{equation}\\

%%%%%%%%%%%%%%%%%%%%%%%%%%%%%%%%%%%%%%%%%%%%%%%
 After diagonalization of the mass matrices ${\bf M}_q$, one obtains the mixing matrix ${\bf V}$. For Hermitian mass matrices the quark mixing matrix ${\bf V}$ can be written as 
\begin{equation}\label{eq:I.7}
{\bf V}={{\bf O}_{u}}^{T}{\bf P}^{u-d}{\bf O}_{d}, 
\end{equation}
where ${\bf P}^{u-d}$ is the diagonal matrix of relative phases,
\begin{equation}\label{eq:I.9}
{\bf P}^{u-d}=diag(e^{i(\phi^u_1-\phi^d_1)},~e^{i(\phi^u_2-\phi^d_2)},~e^{i(\phi^u_3-\phi^d_3)}).
\end{equation}
 Since in the Standard Model there is a single source of CP violation, only one phase difference in Eq. (\ref{eq:I.9}) is observable. Using the freedom in choosing the unobservable phases of the quark fields, it is always possible to take $\phi^u_1=\phi^d_1$. Without loss of generality we can choose the diagonal matrix given in Eq. (\ref{eq:I.9}), with the relative phases as follows, 
\begin{equation}\label{eq:1.23}
\phi^{u}_1-\phi^{d}_1=0,\quad\quad\phi^{u}_2-\phi^{d}_2=\phi^{u}_3-\phi^{d}_3= \Phi.
\end{equation}
Hence, the quark mixing matrix ${\bf V}$ is a function of only one free CP violating phase $\Phi$.\\
We can compute the commutator of the quark mass matrices directly from Eqs. (\ref{eq:I.3})-(\ref{eq:1.7}) and using the phase relations given in Eqs. (\ref{eq:1.14}) and (\ref{eq:1.23}), it is easy to see that the Jarlskog invariant is given as;

\begin{equation}\label{eq:1.31}
J=\frac{T_1\sin\Phi+T_2\sin(2\Phi)}{2F},
\end{equation}
here $T_1$ and $T_2$ depend only on the mass matrix elements given in Eq. (\ref{eq:1.11}); the strength of CP violation depends on $\Phi$. 

In most parameterizations \cite{ref:3} the Jarlskog invariant $J$ is proportional to $\sin\Phi$. In consequence, there is CP violation if the phase $\Phi$ is different from zero or $180^{\circ}$. It is clear that $J$ will have a maximum or a minimum value if the first derivative of $J$ with respect to $\Phi$ is zero, this implies $\Phi=\Phi^*=90^{\circ}$. Furthermore, from Eqs. (\ref{eq:I.7})-(\ref{eq:1.23}) and computing the first and second derivatives of ${\bf V}_{ij}{\bf V}^*_{ij}$ with respect to $\Phi$ one obtains that all of the squared quark mixing moduli ${\bf V}_{ij}{\bf V}^*_{ij}$ have an inflection point for $\Phi=\Phi^*=90^{\circ}$, that is, democratic CP violation is realized in Nature. 
In this way, the fact that there is only one CP violating phase $\Phi$ implies that if $J$ takes a maximum or a minimum value for one particular value of the phase $\Phi$, $\Phi^*=90^{\circ}$, then the quark mixing matrix moduli entries should also take a special value for this particular value of $\Phi^*$. The hypothesis of maximal Jarlskog invariant and democratic CP violation fixes the CP violating phase $\Phi$ to $\Phi=\Phi^*=90^{\circ}$.\\
Pushing forward this idea, we are naturally led to Hermitian quark mass matrix textures which allow maximal CP violation and maximal Jarlskog invariant $J$.
 We find phenomenologically viable mass matrix textures which yield a maximum for the Jarlskog invariant $J$ when the phase $\Phi$ takes the value $\Phi^*=90^{\circ}$.\\
Flavor permutational symmetry of the standard model and an assumed symmetry breaking pattern from which the phenomenologically allowed Hermitian quark mass matrix textures are derived can be the theoretical origin of the quark mixing matrix phase $\Phi^*$.\\ 

This paper is organized as follows: In section \ref{sec:II}, we find the conditions to obtain the Hermitian quark mass textures which give maximal Jarlskog invariant $J$ and maximal CP violating phase $\Phi$. 
 In section \ref{sec:III} we show that all Hermitian mass matrices can be obtained from flavor permutational symmetry breaking.
In section \ref{sec:V} we study the mass patterns derived from maximal Jarlskog invariant and maximal CP violation. We determine that four of the so-called Ramond-Roberts-Ross (RRR) \cite{ref:39} patterns are derived from these conditions. In section \ref{sec:VI}, we undertake a chi square $(\chi^2)$ fit of the quark mixing matrix to show that maximal CP violation is phenomenologically viable.
 Finally, in section \ref{h3} we give our conclusions.

\section{Hermitian mass matrices and maximal CP violation }
\label{sec:II}
 The Jarlskog invariant $J$ takes its maximum or minimum value when the first derivative of $J$ with respect to the CP violating phase $\Phi$ vanishes;

\begin{equation}\label{eq:1.37}
\frac{\partial J}{\partial \Phi}{\bigg |}_{\Phi^*}= 0.
\end{equation}
In general, the maximum or minimum value of $J$ is not explicitly exhibited if we compute the first and second derivatives of $J$ with respect to the CP violating phase $\Phi$.

In order to have maximal Jarlskog invariant $J$, we demand that the first derivative of the Jarlskog invariant $J$ with respect to the CP violating phase $\Phi$ vanishes for $\Phi=\Phi^*$, that is;
\begin{equation}\label{eq:1.39}
T_1\cos\Phi^*+2T_2\cos(2\Phi^*)=0
\end{equation}
Three solutions for Eq. (\ref{eq:1.39}) are interesting:
\begin{itemize}
\item
$\Phi^*=0,~\pi$ and $T_1=-2T_2$; in this case $J$ has an inflection point, CP is an exact symmetry and the Jarlskog invariant $J$ is zero. That is, the Jarlskog invariant $J$ as a function of the CP violating phase $\Phi$ does not take a maximal or minimal value for $\Phi^*=0^{\circ},~180^{\circ}$.
\item
$\Phi^*=45^{\circ},~135^{\circ}$ and $T_1=0$; this solution corresponds to maximal Jarlskog invariant $J$ and non-maximal CP violating phase $\Phi$.
\item
$\Phi^*=90^{\circ},~270^{\circ}$ and $T_2=0$; this solution corresponds to maximal Jarlskog invariant $J$ and maximal CP violating phase $\Phi$.
\end{itemize}
We can satisfy the solutions of Eq. (\ref{eq:1.39}) by taking particular values for the Hermitian quark mass matrix entries and look for the phenomenological implications of the exact zeros.\\
This means that the basis in which the Jarlskog invariant takes its maximum or minimum value is a basis in the space of flavors in which the mass matrices exhibit exact texture zeros for both up and down type quark sectors. We can study  the implications of maximal CP violation in the quark mixing matrix.

\subsection{The quark mixing matrix}

In the mass basis, CP violation is related to the quark mixing matrix, this implies that the quark mixing matrix is also a function of the CP violating phase $\Phi$. We can always write the diagonal phase matrix ${\bf P}^{u-d}$ given in Eq. (\ref{eq:I.9}) as 
\begin{equation}\label{eq:1.41}
{\bf P}^{u-d}={\bf P}_1+{\bf P}_2e^{i\Phi},
\end{equation}
where ${\bf P}_1$ and ${\bf P}_2$ are diagonal matrices. The quark mixing matrix can be written as 

\begin{equation}\label{eq:1.43}
{\bf V}={\bf G}+{\bf G'}e^{i\Phi}, 
\end{equation}
where 
\begin{equation}\label{eq:1.45}
{\bf G}={{\bf O}_{u}}^{T}{\bf P}_1{\bf O}_{d},\qquad\qquad{\bf G'}={{\bf O}_{u}}^{T}{\bf P}_2{\bf O}_{d}.
\end{equation}
Only the second term on the right hand side of Eq. (\ref{eq:1.43}) depends on the CP violation phase $\Phi$. When the flavor mixing matrix ${\bf V}$ is written as in Eq. (\ref{eq:1.43}) the complex phase $\Phi$ describing CP violation appears in all elements of ${\bf V}$. 

The invariant measures of the quark mixing matrix are the moduli of its elements, i.e. the quantities ${\bf V}^{*}_{ij}{\bf V}_{ij}$, the Jarlskog invariant $J$ and the inner angles of the unitarity triangles. Computing the derivative of the quark mixing matrix elements given in Eq. (\ref{eq:1.43} with respect to the CP violation phase $\Phi$, we obtain
\begin{equation}\label{eq:1.47}
\frac{\partial{\bf V}^{\dagger}_{ij}{\bf V}_{kl}}{\partial\Phi}=i\left({\bf G}_{ij}{\bf G'}_{kl}e^{i\Phi}-{\bf G'}_{ij}{\bf G}_{kl}e^{-i\Phi}\right).
\end{equation}
For the second derivative we have 
\begin{equation}\label{eq:1.49}
\frac{\partial^2{\bf V}^{*}_{ij}{\bf V}_{kl}}{\partial\Phi^2}=-\left({\bf G}_{ij}{\bf G'}_{kl}e^{i\Phi}+{\bf G'}_{ij}{\bf G}_{kl}e^{-i\Phi}\right). 
\end{equation}
From Eqs. (\ref{eq:1.47}) and (\ref{eq:1.49}) we have for i=k, j=l three non trivial cases 
\begin{itemize}
\item
When the CP violating phase takes the values $\Phi^*=0^{\circ},~180^{\circ}$, CP is an exact symmetry, all of the elements ${\bf V}_{ij}$ of the quark mixing matrix are real and all of the moduli of the quark mixing matrix $|{\bf V}_{ij}|^2$ take a minimum or a maximum value for $\Phi^*=0^{\circ},~180^{\circ}$.
\item
When the CP violating phase takes the value $\Phi^*=45^{\circ},~135^{\circ}$ the first and second derivatives of ${\bf V}^{*}_{ij}{\bf V}_{ij}$ are equal, but it does not have any further implications.
\item
When the CP violating phase takes the values $\Phi^*=90^{\circ},~270^{\circ}$ all of the elements of the quark mixing matrix moduli $|{\bf V}_{ij}|^2$ have an inflection point. This is a very general property and implies that none of the quark mixing matrix moduli are preferred by Nature, in consequence none of the quark mixing matrix moduli entries are maximal or minimal when the CP violation phase $\Phi$ takes its maximal value, that is $\Phi^*=90^{\circ}$.
\end{itemize}
In a theory with Hermitian mass matrices and only one CP violating phase, all of the quark mixing matrix moduli elements $|{\bf V}_{ij}|^2$ have an inflection point for $\Phi^*=90^{\circ},~270^{\circ}$. In this picture of the quark mixing matrix the magnitude of the CP violating phase $\Phi$ can be fixed demanding that the Jarlskog invariant $J$ takes its maximum value. This gives the solution with $\Phi^*=90^{\circ},~270^{\circ}$ and $T_2=0$ and means that the basis in which the Jarlskog invariant $J$ takes its maximum value is a basis in the space of flavors in which $T_2=0$ imposes exact zeros for the mass matrices, the CP violating phase $\Phi$ takes its maximal value and all of the quark mixing matrix moduli elements have an inflection point. This particular property of the quark mixing matrix ${\bf V}$ and the Jarlskog invariant $J$ is the main result of this paper.\\

%%%%%%%%%%%%%%%%%%%%%%%%%%%%%%%%%%%%%%%%%%%%%%%%%%%%%%%%%%%%%%%%%%%%%%%%%%%%
\section{Hermitian matrices from the breaking of $S_{L}(3)\otimes S_{R}(3)$ }
\label{sec:III}
 In the Standard Model, prior to the introduction of the
Higgs boson and mass terms, the Lagrangian is chiral and invariant
with respect to any permutation of the left and right quark fields, as a consequence of this the left and right quark fields are transformed independently. That is \cite{ref:1}
\begin{eqnarray}\label{eq:3.3}
\Psi^q_L(x)\rightarrow\Psi^{'q}_L(x) ={\bf g}\left(
\matrix{
\psi^q_{1L}(x)\cr
\psi^q_{2L}(x)\cr
\psi^q_{3L}(x)\cr
}\right),
\qquad\qquad
\Psi^q_R(x)\rightarrow \Psi^{'q}_R(x)={\tilde{\bf g}}\left(
\matrix{
\psi^q_{1R}(x)\cr
\psi^q_{2R}(x)\cr
\psi^q_{3R}(x)\cr
}\right)
\end{eqnarray}
where ${\bf g}~\epsilon ~S_L(3)$ transforms the left fields, and $\tilde{\bf g}~\epsilon ~S_R(3)$ transforms the right fields. In this expression, ${\Psi}^d_{L,R}(x)$ and ${\Psi}^u_{L,R}(x)$ denote the left and right quark $d$- and $u$-fields in the current or weak basis respectively, ${\Psi}^{q}(x)$ is a column matrix, its components ${\psi}^q_{k}(x)$ are the quark Dirac fields, $k$ is the flavor index.\\

In this way, the flavor permutational symmetry group of the bilinear form, 
\begin{equation}\label{eq:3.1}
L_{y}=\sum_{i=1}^{3}\sum_{j=1}^{3}\left(\bar{\psi}^u_{Li}\Gamma^u_{ij}\phi\psi^u_{Rj}+\bar{\psi}^d_{Rj}\Gamma^{d}_{ij}\phi\psi^d_{Li}\right)+ h.c.,
\end{equation}
 is $S_L(3)\otimes S_R(3)$, with elements all the pairs $({\bf g},~\tilde{\bf g})$ with ${\bf g}~\epsilon~S_L(3)$ and $\tilde{\bf g}~\epsilon~S_R(3)$.\\
The charged currents are transformed under the flavor group $S_L(3)\otimes S_R(3)$ as:
\begin{eqnarray}\label{eq:3.9}
J^{\pm}_{\mu}\rightarrow  J^{'\pm}_{\mu}=\bar{\psi}^{u}_{L}{\bf g}^{-1}_u{\bf g}_d\gamma_{\mu}{\psi}^{d}_{L}+h.c..
\end{eqnarray}
From this expression it is clear that the charged currents $J^{\pm}_{\mu}$ are invariant under the transformations of the flavor symmetry group if and only if ${\bf g}_u$ and ${\bf g}_d$ are the same matrix.

 Thus, the charged current invariance condition under the family symmetry group implies that the up and down quark fields are transformed with the same group.\\

When the gauge symmetry is spontaneously broken, the quarks and leptons acquire mass and the chiral symmetry of the theory is broken. Then left and right fields are not independent.
The quark mass terms in the Lagrangian gives rise to quark mass matrices ${\bf M}_{d,W}$ and ${\bf M}_{u,W}$,

\begin{equation}\label{eq:1.3}
{\cal L}_{Y}={\bar{\Psi}}^d_{L}{\bf M}_{d,W}{\Psi}^d_{R}+
{\bar{\Psi}}^u_{L}{\bf M}_{u,W}{\Psi}^u_{R}+h.c.
\end{equation}

%%%%%%%%%%%%%%%%%%%%%%%%%%%%%%%%%%%%%%%%%%%%%%%%%%%%%%%%%%%%%%%%%
 We assume that under the flavor symmetry group the quark fields are transformed in the following way:
\begin{eqnarray}\label{eq:3.13}
\bar\Psi^q(x)\rightarrow\bar\Psi^{'q}(x)&=&{{\bf g}\left(\matrix{
\psi^q_{1L}(x)\cr
\psi^q_{2L}(x)\cr
\psi^q_{3L}(x)\cr
}\right) }+{\bf g}{\left(\matrix{
\psi^q_{1R}(x)\cr
\psi^q_{2R}(x)\cr
\psi^q_{3R}(x)\cr
}\right) }.
\end{eqnarray}
The left and right components of the same field are transformed with the same group element. The flavor symmetry group of the bilinear form (\ref{eq:1.3}) is the group $S^{diag.}(3)$ whose elements are the pairs $({\bf g}, ~{\bf g}')$ with first element ${\bf g}~\epsilon~S_L(3)$ and second element ${\bf g}'~\epsilon~S_R(3)$, and ${\bf g}={\bf g}'$. Clearly, $S^{diag.}(3)~\subset S_L(3)\otimes S_R(3)$.\\
The mass term in the Yukawa coupling (\ref{eq:1.3}), transforms under the family symmetry group $S_L(3)\otimes S_R(3)$, in the following way:

\begin{eqnarray}\label{eq:3.17}
L_{y}\rightarrow L'_{y}=\bar{\psi}^{u}_{L}{\bf g}^T{\bf M}_{u,W}{\bf g}\psi^{u}_{R}+\bar{\psi}^{d}_{R}{\bf g}^{T}{\bf M}_{d,W}{\bf g}\psi^{d}_{L}+h.c.
\end{eqnarray}
From here we obtain that under the flavor symmetry group $S^{diag}(3)$, the mass matrices ${\bf M}^q$ are transformed according to the rule:
\begin{eqnarray}\label{eq:3.19}
{\bf M}'_{u,W}={\bf g}^T{\bf M}_{u,W}{\bf g}\qquad and\qquad{\bf M}'_{d,W}={\bf g}^{T}{\bf M}_{d,W}{\bf g}.
\end{eqnarray}
 The Yukawa sector is invariant under the flavor family group $S^{diag}(3)$ if,
\begin{eqnarray}\label{eq:3.21}
{\bf M}'_{q,W}={\bf M}_{q,W}.
\end{eqnarray}
Thus, if the mass matrix ${\bf M}_{q,W}$ commutes with all the elements of the flavor group $S_3$, the Yukawa sector of the Standard Model will have this symmetry.

%%%%%%%%%%%%%%%%%%%%%%%%%%%%%%%%%%%%%%%%%%%%%%%%%%%%%%%%%%%%%%%%%%%%%%%%%%%%%%%%%%
\subsection{Flavor symmetry breaking}
\label{sec:III.1}
 A number of authors \cite{ref:1}-\cite{ref:23} have pointed out that realistic quark mass matrices result from the flavor permutational symmetry $S_{L}(3)\otimes S_{R}(3)$ and its spontaneous or explicit breaking. As an attempt to provide a realistic and general symmetry breaking pattern which clarifies the possible origin of the texture zeros in the Hermitian mass matrices, we suggest an $S(3)_{diag}$ matrix Ansatz and its spontaneous or explicit breaking.
Under exact $S_{L}(3)\otimes S_{R}(3)$ symmetry, the mass spectrum, for either up or down quark sectors, consists of one massive particle (top and bottom quarks) in a singlet irreducible representation and a pair of massless particles in a doublet irreducible representation.
The group $S(3)$ treats three objects symmetrically, while the
hierarchical nature of the mass matrices is a consequence of the
representation structure $\bf{1\oplus2}$ of $S(3)$, which treats the
generations differently. The group $S(3)$ is non-Abelian and has a doublet $|V_1>$ and $|V_2>$ and a singlet $|V_3>$,
\begin{equation}
\label{eq:3.25}
|V_1> = \frac{1}{\sqrt{2}}\pmatrix{ 1 \cr -1 \cr 0 \cr}, \quad 
|V_2>  = \frac{1}{\sqrt{6}}\pmatrix{ 1 \cr 1 \cr -2 \cr}, \quad 
|V_3>  = \frac{1}{\sqrt{3}}\pmatrix{ 1 \cr 1 \cr 1 \cr}.  
\end{equation}
The most general $3\times 3$ hermitian mass matrix ${\bf M}_{q,W}$ contains 9 real free independent parameters, namely, $ C_q,~D_q,~E_q,~A_{1q},~A_{2q},~F_{1q},~F_{2q},~E_{1q}$ and $E_{2q}$. These free parameters emerge in different stages of the flavor permutational symmetry breaking.\\
In the weak basis, the mass matrix with the exact
 $S_{L}(3)\otimes S_{R}(3)$ symmetry reads
\begin{equation}\label{eq:3.27}
{\bf M}_{3q,W}= m_{3q}C_q|V_3> <V_3|,
\end{equation}
where $m_{3q}$ denotes the mass of the third family quark, ${\it t}$ or
${\it b}$. 
To generate masses for the second family, one has to break the
permutational symmetry $S_{L}(3)\otimes S_{R}(3)$ down to
$S_{L}(2)\otimes S_{R}(2)$. This may be done by adding to
${\bf M'}_{3q,W}$ a term ${\bf M}_{2q,W}$ which is invariant under $S_{L}(2)\otimes S_R(2)$ but breaks $S_{L}(3)\otimes S_R(3)$. This can be done with two different matrices, with well defined symmetry properties. In the first case we can assume that ${\bf M}_{2q}$ transforms as the tensorial representation of the vector $|V_2>$,

\begin{equation}\label{eq:3.29}
{\bf M}^2_{2q,W}=m_{3q}D_q|V_2> <V_2|.
\end{equation}
In the second case, we can assume that ${\bf M}_{2q}$ transforms as the complex tensorial symmetry term that mixes the singlet $|V_3>$ with the doublet vector $|V_2>$. Then, in the weak basis, ${\bf M}_{2q}$ is given by
\begin{equation}\label{eq:3.31}
{\bf M}^3_{2q,W}=m_{3q}\left[B_{1q}\left(|V_3> <V_2|+|V_2> <V_3|\right)+iB_{2q}\left(|V_3> <V_2|-|V_2> <V_3|\right)\right].
\end{equation}
We may now turn our attention to the question of breaking the
$S_L(2)\otimes S_R(2)$ symmetry. In order to give mass to the first family, we add another term ${\bf M}_{1q}$ to the mass matrix. This breaking can be obtained in three different ways:  First assume that ${\bf M}_{1q}$ transforms as the tensorial representation of the vector $|V_1>$;

\begin{equation}\label{eq:3.33}
{\bf M}^1_{1q,W}=m_{3q}E_q|V_1> <V_1|.
\end{equation}
Putting the first family in a complex representation will allow us to have a CP violating phase in the mixing matrix. This breaking can be done in two different ways, we can assume that ${\bf M}_{1q}$ transforms as the mixed symmetry term of the doublet complex tensorial representation of the $S(3)_{d}$ diagonal subgroup of $S_{L}(3)\otimes S_{R}(3)$. Then, in the weak basis, ${\bf M}_{1q}$ is given by
\begin{equation}\label{eq:3.35}
\begin{array}{c}
{\bf M}^2_{1q,W}=m_{3q}\left[ A_{1q}\left(|V_1> <V_2|+|V_2> <V_1|\right)+iA_{2q}\left(|V_1> <V_2|-|V_2> <V_1|\right)\right]
\end{array}
\end{equation}
We can assume that ${\bf M}_{1q}$ transforms as the complex tensorial symmetry term that mixes the singlet $|V_3>$ with the doublet vector $|V_1>$. Then, in the weak basis, ${\bf M}_{1q}$ is given by
\begin{equation}\label{eq:3.37}
\begin{array}{c}
{\bf M}^3_{1q,W}= m_{3q}\left[ F_{1q}\left(|V_3> <V_1|+|V_1> <V_3|\right)+iF_{2q}\left(|V_1> <V_3|-|V_3> <V_1|\right)\right]
\end{array}
\end{equation}
Finally, adding the quark mass matrices given in Eqs. (\ref{eq:3.27}-\ref{eq:3.37}), we get the mass matrix ${\bf M}_{q,W}$ in the weak basis.
%%%%%%%%%%%%%%%%%%%%%%%%%%%%%%%%%%%%%%%%%%%%%%%%%%%%%%%%%%%%%%%%%%
%%%%%%%%%%%%%%%%%%%%%%%%%%%%%%%%%%%%%%%%%%%%%%%%%%%%%%%%%%%%%%%%%%
\subsection{Mass matrix from flavor symmetry breaking}\label{sec:IV}

To make explicit the assignment of particles to irreducible
representations of $S_{L}(3)\otimes S_{R}(3)$, it will be convenient
to make a change of basis from the weak basis to a 
hierarchical basis. In this basis, the quark fields are

\begin{equation}
\label{eq:4.1}
\psi_{1q,H}(x) = \frac{1}{\sqrt{2}} (\psi_{1q,W}(x) - \psi_{2q,W}(x)),
\end{equation}

\begin{equation}
\label{eq:4.3}
{\psi_{2q,H}(x) = \frac{1}{\sqrt{6}} (\psi_{1q,W}(x) + \psi_{2q,W}(x) -
    2\psi_{3q,W}(x) )},
\end{equation}

\begin{equation}
\label{eq:4.5}
{\psi_{3q,H}(x) = \frac{1}{\sqrt{3}} (\psi_{1q,W}(x) + \psi_{2q,W}(x) + \psi_{3q,W}(x)
  )},
\end{equation}

\noindent
the subindex $H$ denotes the hierarchical basis. In the hierarchical basis, the third family quarks, $t$ or $b$, are assigned to the invariant singlet irreducible representation $\Psi_{3q,H}(x)$, the other two families are assigned to $\Psi_{2q,H}(x)$ and $\Psi_{1q,H}(x)$, the two components of the doublet irreducible representation of $S_{diag}(3)$.

The mass matrix ${\bf M}_{q,H}$ in the hierarchical
 basis is related to the mass matrix in the weak basis by the
unitary transformation

\begin{equation}
\label{eq:4.7}
{\bf  M_{q,H}} = {\bf U}^{\dagger}{\bf M_{q,W}}{\bf U},
\end{equation}
where
\begin{equation}
\label{eq:4.9}
{\bf U}={1\over \sqrt {6}}\pmatrix{
\sqrt {3} & 1 & \sqrt {2} \cr
-\sqrt {3} & 1 & \sqrt {2} \cr
0 & -2 & \sqrt {2} \cr
}.
\end{equation}
Then, in the hierarchical basis, the mass matrix ${\bf M}_{q,W}$ takes the form given in Eq. (\ref{eq:1.7}) with,
\begin{equation}\label{eq:eq:4.13}
\begin{array}{rcl}
A_q=\sqrt{A^2_{1q}+A^2_{2q}},\qquad B_q=\sqrt{B^2_{1q}+B^2_{2q}},\qquad F_q=\sqrt{F^2_{1q}+F^2_{2q}},
\end{array}
\end{equation}
and the phases are given by the following relations 
\begin{equation}\label{eq:eq:4.15}
\begin{array}{rcl}
 \phi^q_A=\tan^{-1}\left(\frac{A_{2q}}{A_{1q}}\right),\qquad \phi^q_B=\tan^{-1}\left(\frac{B_{2q}}{B_{1q}}\right),\qquad \phi^q_F=\tan^{-1}\left(\frac{F_{2q}}{F_{1q}}\right),
\end{array}
\end{equation}
which are also given in Eq. (\ref{eq:1.14}).\\

At this stage in our derivation of Hermitian mass matrices a question comes naturally to mind: Which flavor permutational symmetry breaking patterns are realized in Nature? or phrased differently: Which of the Hermitian mass matrix textures are phenomenologically allowed?.
To answer this question, we will analyze the phenomenological implications of maximal Jarlskog invariant $J$ and maximal CP violating phase $\Phi^*=90^{\circ}$.\\
%%%%%%%%%%%%%%%%%%%%%%%%%%%%%%%%%%%%%%%%%%%%%%%%%%%%%%%%%%%%%%%
\section{Mass textures from maximal Jarlskog invariant $J$ }
\label{sec:V}
We are now in a position to consider the realistic $3\times 3$ Hermitian mass matrices which allow maximal Jarlskog invariant $J$ and maximal CP violating phase $\Phi$. In order to have maximal CP violation, we demand that the first derivative of the Jarlskog invariant $J$ with respect to the CP violating phase $\Phi$ vanishes
for $\Phi^*=90^{\circ}$, that is;
\begin{equation}\label{eq:5.3}
T_2=0
\end{equation}
where 
\begin{eqnarray}\label{eq:5.4}
\frac{T_2}{2}&=& F_u A_d [A_u (D_d-E_d)+F_u B_d][F_d(C_u-E_u)+B_u A_d] \cr &+&
F_d A_u[F_u (C_d-E_d)+A_u B_d][ A_d(D_u-E_u)+B_u F_d]\cr&+&2A_uA_d[F_u(C_d- E_d)+A_u B_d][F_d(C_u-E_u)+A_d B_u] \cr&+& 2 F_uF_d[A_u(D_d-E_d)+F_uB_d][A_d(D_u-E_u)+F_d B_u]\cr&+&(A_uF_d-A_dF_u)(A_uA_d+F_uF_d)[B_d(C_u-D_u)+B_u(D_d-C_d)]
\end{eqnarray}
We can satisfy Eq. (\ref{eq:5.3}) by taking particular values for the Hermitian quark mass matrix entries and look for the phenomenological implications of these exact zeros. First of all notice that the quark mass matrices ${\bf M}_{u}$ and ${\bf M}_{d}$ commute when 
\begin{equation}\label{eq:5.5}
A_u=F_u=0\qquad\qquad or\qquad\qquad A_d=F_d=0
\end{equation}
this implies that the Jarlskog invariant $J$ is zero for these textures and CP violation is related to the condition 
\begin{equation}\label{eq:5.7}
A_q\neq 0\qquad or\quad F_q\neq 0\qquad for\qquad $q=u,~d$.
\end{equation}
Thus, CP violation is related to the breaking of the $S_L(2)\otimes S_R(2)$ by the term ${\bf M}^2_{1q,W}$ that transforms as the mixed symmetry term of the doublet complex tensorial representation of the $S(3)_{d}$ diagonal subgroup of $S_{L}(3)\otimes S_{R}(3)$ or to the breaking of the $S_L(2)\otimes S_R(2)$ by the complex tensorial symmetry term ${\bf M}^3_{1q,W}$ that mixes the singlet $|V_3>$ with the doublet vector $|V_1>$.\\

Mass matrices with textures are over-constrained leading to the prediction of relations between mixing angles and quark masses and it is our task to find which, if any, of these textures is viable and compatible with maximal Jarlskog invariant $J$ and maximal CP violating phase $\Phi$. For the assumptions made above, we found 14 texture patterns that satisfy Eq. (\ref{eq:5.3}),

\begin{enumerate}
\item
If we take $F_u=F_d=0$, this gives two solutions $B_u=0$ or $B_d=0$. For the first solution we have the textures
\begin{eqnarray}\label{eq:5.9}
{\bf M}_u=\pmatrix{
E_u & A_u & 0 \cr
A_u & D_u & 0 \cr
0 & 0 & C_u \cr
},\qquad\qquad
{\bf M}_d=\pmatrix{
E_d & A_d & 0 \cr
A_d & D_d & B_d \cr
0 & B_d & C_d \cr
},
\end{eqnarray}
and the second solution is obtained if we change $u\leftrightarrow d$ in Eq. (\ref{eq:5.9}).
\item
If we take $A_u=A_d=0$, this gives two solutions $B_u=0$ or $B_d=0$. For the first solution we have the textures
\begin{eqnarray}\label{eq:5.13}
{\bf M}_u=\pmatrix{
E_u & 0 & F_u \cr
0 & D_u & 0 \cr
F_u & 0 & C_u \cr
},\qquad\qquad
{\bf M}_d=\pmatrix{
E_d & 0 & F_d \cr
0 & D_d & B_d \cr
F_d & B_d & C_d \cr
},
\end{eqnarray}
and the second solution is obtained if we change $u\leftrightarrow d$ in Eq. (\ref{eq:5.13}).

\item
If we take $A_u=F_d=0$, this gives two solutions $B_u=0$ or $B_d=0$. For the first solution we have the textures
\begin{eqnarray}\label{eq:5.17}
{\bf M}_u=\pmatrix{
E_u & 0 & F_u \cr
0 & D_u & 0 \cr
F_u & 0 & C_u \cr
},\qquad\qquad
{\bf M}_d=\pmatrix{
E_d & A_d & 0 \cr
A_d & D_d & B_d \cr
0 & B_d & C_d \cr
};
\end{eqnarray}
for the second solution we have
\begin{eqnarray}\label{eq:5.19}
{\bf M}_u=\pmatrix{
E_u & 0 & F_u \cr
0 & D_u & B_u \cr
F_u & B_u & C_u \cr
},\qquad\qquad
{\bf M}_d=\pmatrix{
E_d & A_d & 0 \cr
A_d & D_d & 0 \cr
0 & 0 & C_d \cr
}.
\end{eqnarray}

\item
If we take $A_d=F_u=0$, this gives two solutions $B_u=0$ or $B_d=0$. These textures can be obtained by changing $u\leftrightarrow d$ in the textures given in Eqs. (\ref{eq:5.19}) and (\ref{eq:5.17}) respectively.

\item
Taking $B_u=B_d=0$, we obtain six solutions, $F_q=0$ or $A_q=0$ or  $E_q=D_q=C_q$ where $q=u,~d$. The textures for $q=u$ are the following;

\begin{eqnarray}\label{eq:5.15}
{\bf M}_u=\pmatrix{
E_u & A_u & 0 \cr
A_u & D_u & 0 \cr
0 & 0 & C_u \cr
},\qquad\qquad
{\bf M}_d=\pmatrix{
E_d & A_d & F_d \cr
A_d & D_d & 0 \cr
F_d & 0 & C_d \cr
};
\end{eqnarray}
for the second solution we have
\begin{eqnarray}\label{eq:5.21}
{\bf M}_u=\pmatrix{
E_u & 0 & F_u \cr
0 & D_u & 0 \cr
F_u & 0 & C_u \cr
},\qquad\qquad
{\bf M}_d=\pmatrix{
E_d & A_d & F_d \cr
A_d & D_d & 0 \cr
F_d & 0 & C_d \cr
};
\end{eqnarray}
for the third solution we obtain,
\begin{eqnarray}\label{eq:5.23}
{\bf M}_u=\pmatrix{
C_u & A_u & F_u \cr
A_u & C_u & 0 \cr
F_u & 0 & C_u \cr
},\qquad\qquad
{\bf M}_d=\pmatrix{
E_d & A_d & F_d \cr
A_d & D_d & 0 \cr
F_d & 0 & C_d \cr
}.
\end{eqnarray}
The solutions for $q=d$ are obtained  by changing $u\leftrightarrow d$ in the textures given in Eqs. (\ref{eq:5.15})-(\ref{eq:5.23}). 
\end{enumerate}

Maximal Jarlskog invariant $J$ and democratic CP violation with maximal CP violating phase $\Phi$ is the underlying theoretical assumption which leads to the structure of the quark mass textures.
This means that the basis in which the Jarlskog invariant $J$ is maximal, and the CP violating phase $\Phi$ is maximal, is a basis in the space of flavors, in which the mass matrices exhibit exact texture zeros for both up and down type quark sectors. 
It is clear from Eqs. (\ref{eq:5.5}-\ref{eq:5.23}) that, the zeros of the quark mass textures ${\bf M}_u$ and ${\bf M}_d$ derived from the condition of maximal Jarlskog invariant and maximal CP violating phase $\Phi$ do not give up the parallelism between the structures of ${\bf M}_u$ and ${\bf M}_d$. \\

Ramond Roberts and Ross (RRR ) \cite{ref:39} have found that there exist five phenomenologically allowed patterns of Hermitian quark mass matrices, which have five texture zeros. 
The five allowed patterns are shown in Table \ref{tab:1}. The question raised by RRR \cite{ref:39}: ``What underlying theory can lead to such structure?'' has an answer in terms of the maximal Jarlskog invariant and the maximal democratic CP violating phase $\Phi$.
The RRR patterns I and III-V are obtained when we take $E_u=E_d=0$ in our textures. The mass pattern II can not be obtained from the above assumptions, for this reason this texture pattern could not be a good candidate for the quark mass matrices in an underlying theory of fermion mass generation.\\
The appearance of the texture  zeros in the mass matrices is generally assumed to arise at some high scale Q where new physics connected with mass generation comes into play, and the comparison of a mass matrix pattern which holds at a large scale Q with experiment is complicated by the evolution of the Yukawa couplings with energy. This evolution can change zeros in a mass matrix at a given scale into small, but non-vanishing, contributions at a different scale. The effect of the renormalization group evolution of the couplings is to obscure possible mass matrix patterns. Furthermore, the renormalization group (RG) running may not preserve the hermeticity of the quark mass textures. \\

We follow the Olechowski-Pokorski  \cite{ref:42} paper for the one loop renormalization group equations for the quark mixing matrix elements $|V_{ij}|$ and the RRR \cite{ref:39} analysis of the renormalization group evolution of the CKM matrix and the masses. Keeping the top and bottom Yukawa couplings only and neglecting thresholds, Olechowski and Pokorski \cite{ref:42} and RRR \cite{ref:39} obtained that the CKM matrix elements evolve as

\begin{eqnarray}\label{eq:5.34}
16\pi^2\frac{d |V_{ij}|}{dt}&=&-\frac{3c}{2}(h^2_t+h^2_b)|V_{ij}|, \quad\quad
ij=(13),~(31),~(23),~(32)\cr\cr
16\pi^2\frac{d |V_{12}|}{dt}&\approx&-\frac{3c~h^2_t}{2}\frac{|V_{31}|^2-|V_{13}|^2}{|V_{12}|},\quad\quad
16\pi^2\frac{d |V_{21}|}{dt}\approx-\frac{3c~h^2_b}{2}\frac{|V_{13}|^2-|V_{31}|^2}{|V_{21}|}.
\end{eqnarray}
Here $t=ln(Q/Q_0)$ where the elements are evaluated at the scale $Q$, $h_t$ and $h_b$ are the Yukawa couplings and $c$ is a constant determined by the couplings of the theory.  From Eq. (\ref{eq:1.43}) and (\ref{eq:5.34}), we obtain the running equation for the CP violating phase $\Phi$;
\begin{eqnarray}\label{eq:5.35}
16\pi^2\frac{d \Phi}{dt}&=&0
\end{eqnarray}
that is, the CP violating phase $\Phi$ do not run with energy. The running of the quark mixing matrix elements $|{\bf V}_{ij}|$ yield the following RG equations for the matrices ${\bf G}$ and ${\bf G'}$ introduced in Eq. (\ref{eq:1.45}):
\begin{eqnarray}\label{eq:5.36}
16\pi^2\frac{d G_{ij}}{dt} &=&-\frac{3c}{2}(h^2_t+h^2_b)G_{ij}, \quad
16\pi^2\frac{d G'_{ij}}{dt}=-\frac{3c}{2}(h^2_t+h^2_b)G'_{ij},~~ij=13,~31,~23,~32\cr\cr
16\pi^2\frac{d G_{12}}{dt}&=&~~16\pi^2\frac{d G'_{12}}{dt}\approx~~0, \quad16\pi^2\frac{d G_{21}}{dt}=~~16\pi^2\frac{d G'_{21}}{dt}\approx~~ 0
\end{eqnarray}

Olechowski and Pokorski \cite{ref:42} also found that the Jarlskog invariant $|J|$ evolves via the equation

\begin{eqnarray}\label{eq:5.37}
16\pi^2\frac{d |J|}{dt}=-3c(h^2_t+h^2_b)|J|,
\end{eqnarray}
where we can choose $|J|$ independent of the parametrization given by Eq. (\ref{eq:1.31}), from which we find
 
\begin{eqnarray}\label{eq:5.39}
16\pi^2\frac{d T_1}{dt}=-3c(h^2_t+h^2_b)T_1, \quad\quad
16\pi^2\frac{d T_2}{dt}=-3c(h^2_t+h^2_b)T_2,
\end{eqnarray}
It follows from this equation that the zeros imposed at some high energy scale Q for the constrains $T_1(Q)=0$ or $T_2(Q)=0$ are protected during the renormalization group evolution of the CKM matrix and the masses. This result with that of Eq. (\ref{eq:5.35}) protects the Hermiticity of the mass textures and sheds some light for the comparison with low energy experiments of the mass matrix patterns derived from maximal Jarlskog invariant $J$ and maximal CP violating phase $\Phi$ which holds at a large scale Q.  We can take the same texture patterns at all energies, the only constraint is to express all the running quark masses at some common energy scale.\\
As noticed by Fritzsch and Xing \cite{ref:26}, with the same input values of quark mass ratios only some of the textures patterns are in good agreement with current experimental data and the 14 textures patterns that we derive cannot simultaneously survive. We find three texture patterns which can be confronted with experiment, pattern I and the patterns III and IV in the RRR classification given in Table \ref{tab:1}. These patterns can be written in terms of the quark mass ratios without introducing any new parameters.\\

%%%%%%%%%%%%%%%%%%%%%%%%%%%%%%%%%%%%%%%%%%%%%%%%%%%%%%%%%%%%%%%%%%%%%%%%%%%%
\section{Numerical results}
\label{sec:VI}
Before proceeding to give the numerical results for the mixing matrix ${\bf V}^{th}$, it will be convenient to stress the following points:
\begin{enumerate}
\item
The CP violation phase $\Phi$ is fixed to $\Phi=90^{\circ}$ from the condition of maximal Jarlskog invariant and democratic CP violation.  
\item
The masses of the lighter quarks are less well determined, while the moduli of the entries in $|V^{exp}_{ij}|$ with the largest errors, namely $|V_{ub}|$ and $|V_{td}|$ are sensitive to changes in the values of the ligther quarks $m_u$ and $m_d$ respectively. The sensitivity of $|V_{ub}|$ and $|V_{td}|$ to changes in  $m_u$ and $m_d$ respectively, is reflected in the shape of the unitarity triangle which changes appreciably when the masses of the lighter quarks change within their uncertainty ranges. 
\item
The moduli of the entries in $|V^{exp}_{ij}|$ with the smallest errors, namely $|V_{ud}|$, $|V_{us}|$, $|V_{cd}|$ and $|V_{cs}|$ are the most sensitive to changes in the values of the lighter quarks $m_d$ and $m_s$. Hence, the quality of the fit of $|V^{th}_{ij}|$ to $|V^{exp}_{ij}|$ for $i,~j~=~1,~2$ is spoiled if relatively large changes in the masses of the lighter quarks $m_d$ and $m_s$ are made. This puts a strong constraint on the allowed values for $m_d/m_s$.
The best simultaneous $\chi^2$ fits of $|V^{th}_{ij}|$ for $i,~j~=~1,~2,~3$ to the experimentally determined quantities are very sensitive to the ratio  $m_d/m_s$ and allows us to constrain $|V^{th}_{td}|$.  
\item
For the purpose of calculating quark mass ratios and computing the mixing matrix, it is convenient to give all quark masses as running masses at some common energy scale \cite{ref:24}, \cite{ref:25}. In the present calculation, following Peccei \cite{ref:24}, Fritzsch \cite{ref:26}, the BaBar book \cite{ref:27} and RRR \cite{ref:39}, we used the values of the running quark masses  evaluated at $\mu=m_t$.  The running quark masses  evaluated at $\mu=m_t$ as given by Fusaoka and Koide \cite{ref:25} are in units of $GeV$;
\begin{eqnarray}\label{eq:5.40}
m_d&=&(4.49\pm 0.64)\times 10^{-3},\qquad 
m_u=(2.23\pm 0.43)\times 10^{-3}, \cr\cr
m_s&=&(8.94\pm 1.25)\times 10^{-2},\qquad 
m_c=(6.46\pm 0.59)\times 10^{-1},\cr\cr
m_b&=&2.85\pm 0.11,\qquad\qquad\qquad 
m_t=171\pm 12.
\end{eqnarray}

\end{enumerate}
%%%%%%%%%%%%%%%%%%%%%%%%%%%%%%%%%%%%%%%%%%%%%%%%%%%%%%%%%%%%%%%%%
From the strong hierarchy in the masses of the quark families, 
$1>> m_{2q}/m_{3q}> m_{1q}/m_{3q}$, we expect $C_{q}$ in ${\bf M}_{q}$ to be very close to unity. Computing the invariants of ${\bf M}_{q}$, $tr {\bf M}_{q}$, $tr {{\bf M}_{q}}^{2}$ and $det {\bf M}_{q}$ and by appropriately choosing the signs of the quark masses, the entries in the mass matrix may be readily expressed in terms of the quark masses. From here, we computed ${\bf V}^{th}_{ij}$ with the exact expressions for ${\bf O}_{u}$ and ${\bf O}_{d}$.\\
We use the MINUIT program to perform a $\chi^2$ fit \cite{MINUIT} of the quark mixing matrix moduli $|{\bf V}^{th}_{ij}|$ to $|{\bf V}^{exp}_{ij}|$, $i,~j~=~1,~2,~3$. In the fit, we fixed $\Phi=90^{\circ}$ and vary the quark mass ratios within the allowed limits computed from Eqs. (\ref{eq:5.40}). Once the minimum of the $\chi^2$ is found, we release the quark mass ratios and let them vary without limits. For three different allowed mass patterns the minimum of the $\chi^2$ fit of the exact expressions of the quark mixing matrix $|{\bf V}^{th}_{ij}|$ to $|{\bf V}^{exp}_{ij}|$, $i,~j~=~1,~2,~3$, is obtained. The results for the one-standard-deviation values of the running quark mass ratios are given in Table \ref{tab:2}.\\

\begin{figure}[tbp]
\begin{center}
\setlength{\unitlength}{0.95cm}
\setlength{\fboxsep}{0cm}
\begin{picture}(14,19)
\put(-1,11.5){$\beta$}
\put(1.5,6.7){${\small m_u/m_c}$}
\put(0,7){\begin{picture}(5,5)\put(-0.9,-0.4){\piccie{6.0cm}{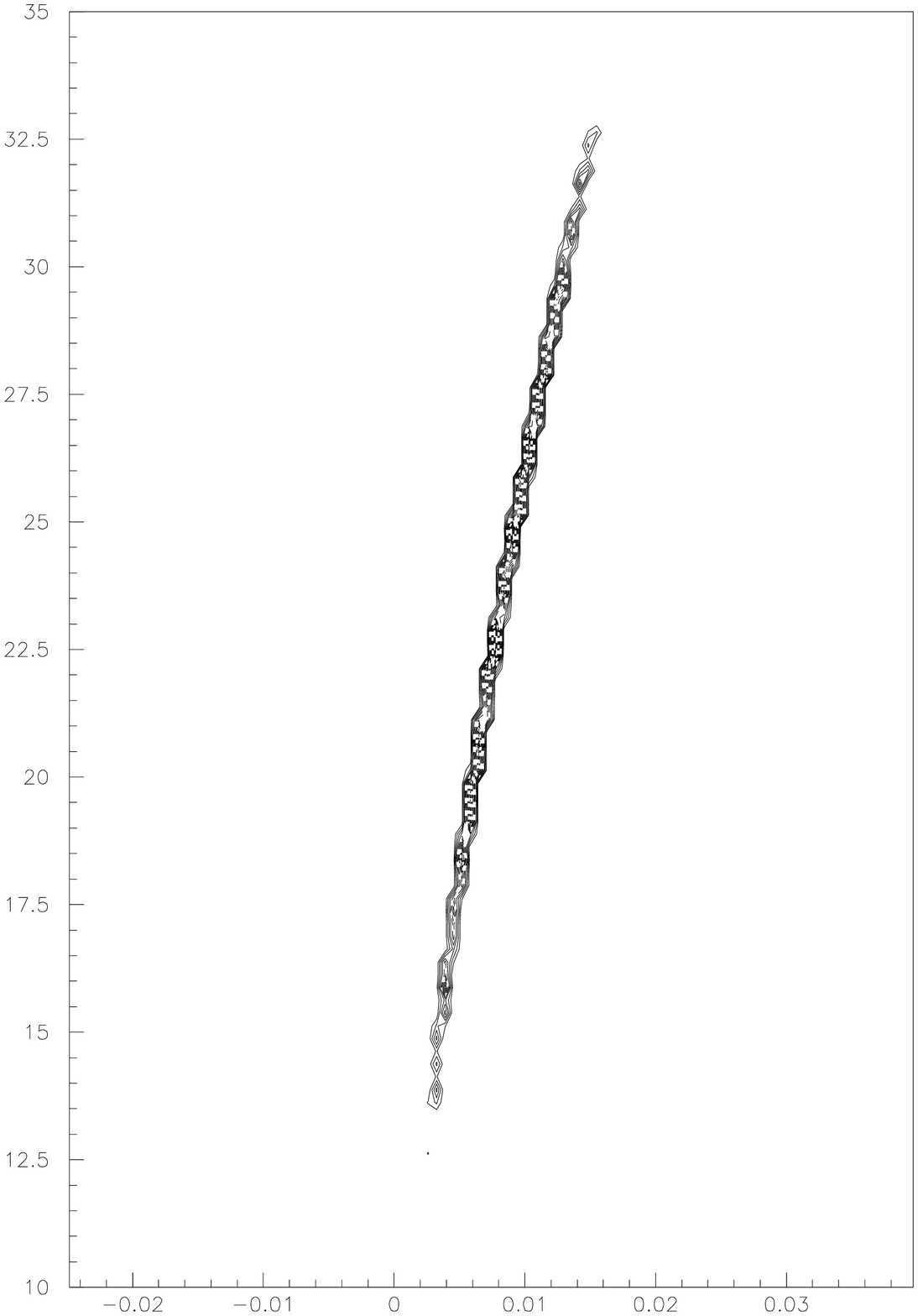}}\end{picture}}
\put(6,11.5){$\beta$}
\put(8.5,6.7){$\alpha$}
\put(7,7){\begin{picture}(5,5)\put(-0.9,-0.4){\piccie{6.0cm}{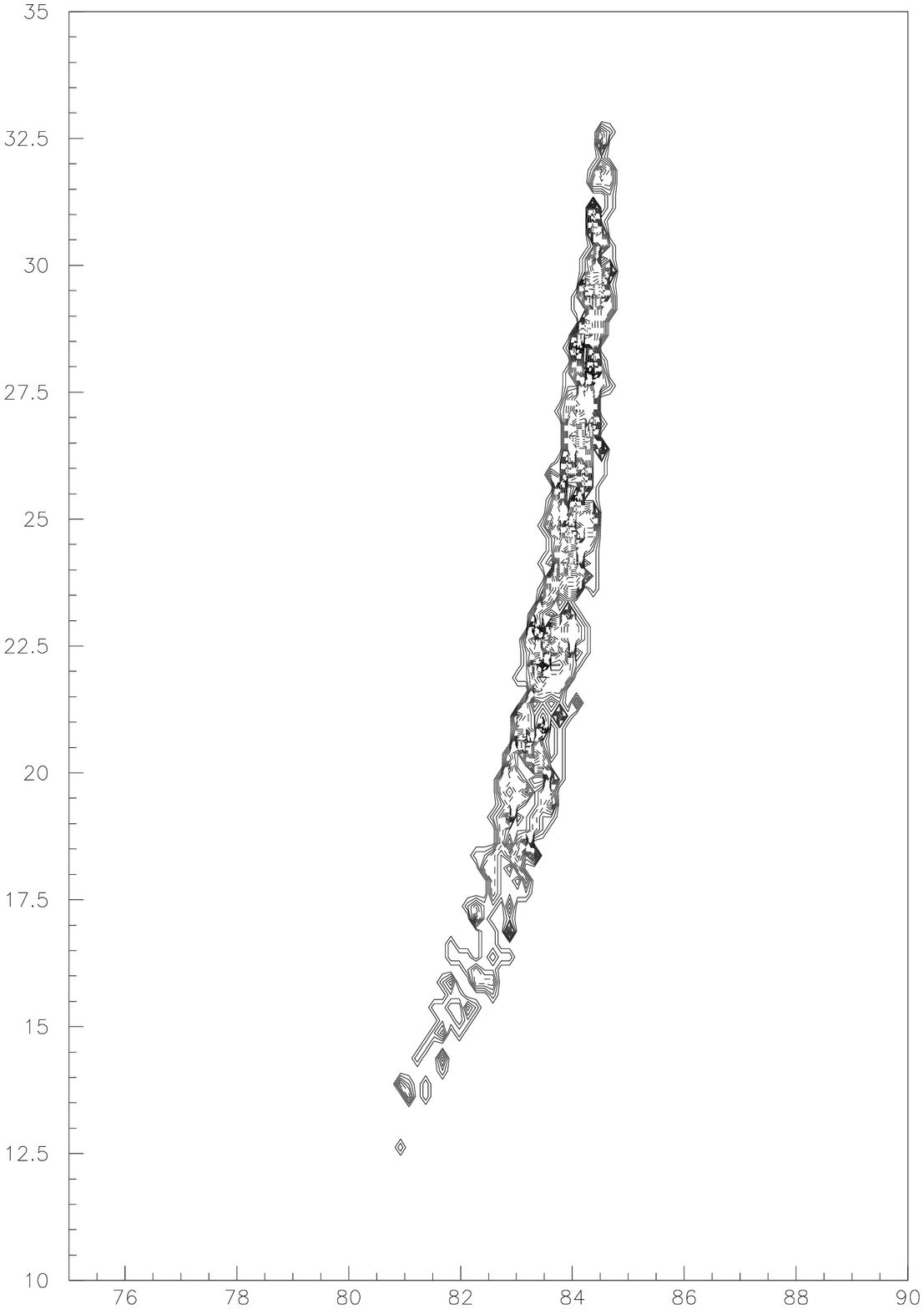}}\end{picture}}
\put(-1,2.5){$\gamma$}
\put(1.5,-2.3){${\alpha}$}
\put(0,-2){\begin{picture}(5,5)\put(-0.9,-0.4){\piccie{6.0cm}{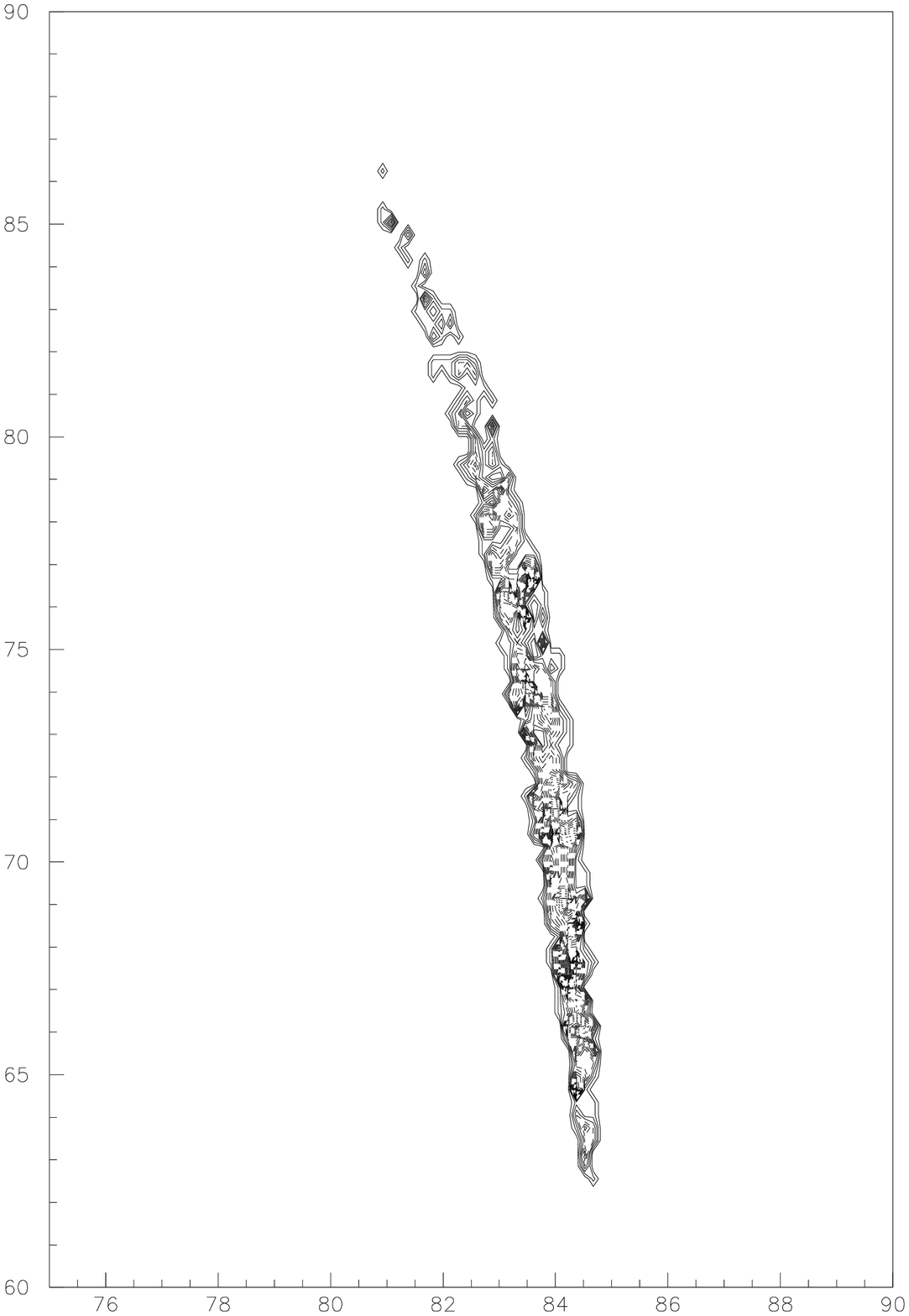}}\end{picture}}
\put(5.6,2.5){${\small \frac{|{\bf V}_{ub}|}{|{\bf V}_{cb}|}}$}
\put(8.5,-2.3){${\small m_u/m_c}$}
\put(7,-2){\begin{picture}(5,5)\put(-0.9,-0.4){\piccie{6.0cm}{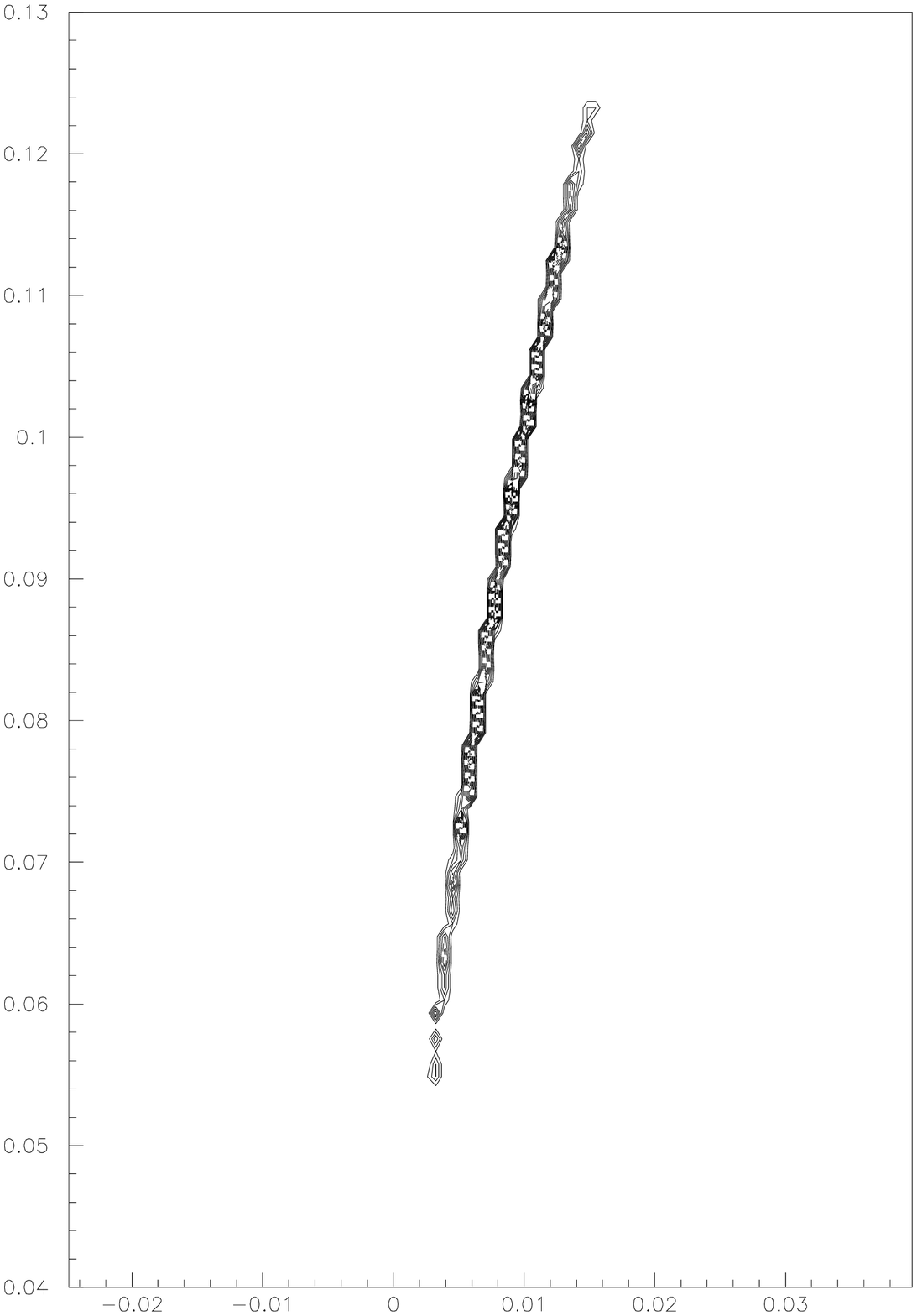}}\end{picture}}
\end{picture}
\end{center}
\vspace{1.cm}\bigskip\bigskip\bigskip
\caption{The one-standard-deviation range of the angles $\beta$, $\gamma$ and the ratio $|V_{ub}|/|V_{cb}|$ computed from the mass pattern I are shown as function of the ratio $mu/mc$. The $\left( \alpha, \beta \right)$ correlation in this mass pattern is also shown.}
\label{fig:1}
\end{figure}

%%%%%%%%%%%%%%%%%%%%%%%%%%%%%%%%%%%%%%%%%%%%%%%%%%%%%%%%%%%%%%%%%%%
For the texture pattern I, the entries $A_u,~ D_u$, and $C_u$ in the mass matrix ${\bar{\bf M}}_u$ and $A_d,~ D_d,~B_d$, and $C_d$ in the mass matrix ${\bar{\bf M}}_d$ may be readily expressed in terms of the quark mass ratios as follows;
\begin{eqnarray}\label{eq:5.56}
{{A}^2_{u}}=\tilde{m}_{u}\tilde{m}_{c},\quad\quad
D_{u}=\tilde{m}_{u}-\tilde{m}_{c},\quad\quad C_u=1,
\end{eqnarray}
\begin{eqnarray}\label{eq:5.43}
{{A}^2_{d}}&=&{\tilde{m}_{d}\tilde{m}_{s} \over{1 -\tilde{m}_{d}}},\quad\quad
D_{d}=2\tilde{m}_{d}-\tilde{m}_{s},\quad\quad C_d=1-\tilde{m}_d,\cr
{{B}^2_{d}}&=&\tilde{m}_d\left[1+2\tilde{m}_{s}\left(1-\frac{1}{2\left(1-\tilde{m}_d\right)}\right)-2\tilde{m}_d\right]
\end{eqnarray}
 The quark mixing matrix for pattern I is a function of three quark mass ratios and $\Phi$, that is, ${\bf V}={\bf V}\left(\frac{mu}{mc},~\frac{md}{mb},~\frac{ms}{mb},~\Phi^*\right)$, the range of values of ${m_d}/{m_b}$  are in the allowed limits computed from  Eqs. (\ref{eq:5.40}) and the quark mass ratio ${m_s}/{m_b}$ is close to the upper bound of the allowed limit as computed from Eqs. (\ref{eq:5.40}), while the high central value of ${m_u}$ obtained from ${m_u}/{m_c}$ agrees with the value of $m_u$ at the top scale found in \cite{ref:39} and \cite{ref:1}. These quark mass ratio values are in good agreement with the phenomenological relations obtained by Fritzsch\cite{ref:26}
\begin{eqnarray}\label{eq:5.58}
\frac{|{\bf V}_{ub}|}{|{\bf V}_{cb}|}\approx\sqrt{\frac{m_u}{m_c}},\qquad \frac{|{\bf V}_{td}|}{|{\bf V}_{ts}|}\approx\sqrt{\frac{m_d}{m_s}},\qquad\sin\theta_c\approx\sqrt{\frac{m_d}{m_s}+\frac{m_u}{m_c}}.
\end{eqnarray}
It is clear from these relations that smaller values of ${m_u}/{m_c}$ favors smaller values of $|{\bf V}_{ub}|/|{\bf V}_{cb}|$ as it is shown in Fig. \ref{fig:1}.\\

%%%%%%%%%%%%%%
 For the  texture pattern III we get 
\begin{eqnarray}\label{eq:5.41}
{{F}^2_{u}}&=&\tilde{m}_{u},\quad\quad
D_{u}=\tilde{m}_{c},\quad\quad C_u=1-\tilde{m}_u,
\end{eqnarray}
The parameters $A_d,~ D_d,~B_d$, and $C_d$ of the quark mass texture ${\bar{\bf M}}_d$, are given by Eq. (\ref{eq:5.43}) in terms of the quark mass ratios.
For pattern III, the quark mixing matrix is a function of the CP violating phase $\Phi$ and three quark mass ratios, that is,
${\bf V}={\bf V}\left(\frac{mu}{mt},~\frac{md}{mb},~\frac{ms}{mb},~\Phi^*\right)$.\\
As in the previous case, in the $\chi^2$ fit of $|{\bf V}^{th}_{ij}|$ to $|{\bf V}^{exp}_{ij}|$ for $i,~j~=~1,~2,~3$ we fixed the CP violation phase $\Phi=90^{\circ}$ and varied the quark mass ratios within the allowed limits computed from Eqs. (\ref{eq:5.40}). Once the minimum is found, we release the quark mass ratios and let them vary without limits.
The minimum of the $\chi^2$ fit of the quark mixing matrix computed from the texture pattern III is obtained for the quark mass ratio values given in Table \ref{tab:2}. The values of the quark mass ratios ${m_u}/{m_t}, ~{m_d}/{m_b}$ and ${m_s}/{m_b}$ obtained for pattern III are close to the central values of the quark mass ratios as obtained from Eqs. (\ref{eq:5.40}). For this pattern the central values of the quark mass ratios gives the central value of $|{\bf V}_{ub}|/|{\bf V}_{cb}|$ which is in very good agreement with its latest determination \cite{ref:3}, \cite{ref:32} and \cite{ref:34}.\\

%%%%%%%%%%%%%%%%%%%%%%%%%%%%%%%%%%%%%%%%%%%%%%%%%%%%%%%%%%%%%%%%%%% 
For the texture pattern IV, the entries $A_u,~ D_u,~B_u$, and $C_u$ in the mass matrix ${\bar{\bf M}}_u$ and $A_d,~ D_d$, and $C_d$ in the mass matrix ${\bar{\bf M}}_d$ are expressed in terms of the quark mass ratios as follows:
\begin{eqnarray}\label{eq:5.73}
{{A}^2_{u}}&=&{\tilde{m}_{u}\tilde{m}_{c} \over{1 -\frac{\tilde{m}_{c}}{2}}},\quad\quad
D_{u}=\tilde{m}_{u}-\frac{\tilde{m}_{c}}{2},\quad\quad C_u=1-\frac{\tilde{m}_c}{2},\cr
{{B}^2_{u}}&=&\frac{\tilde{m}_c}{2}\left[1+\tilde{m}_{u}\left(1-\frac{4}{\left(2-\tilde{m}_c\right)}\right)-\frac{\tilde{m}_c}{2}\right],
\end{eqnarray}
\begin{eqnarray}\label{eq:5.75}
{{A}^2_{d}}=\tilde{m}_{d}\tilde{m}_{s},\quad\quad
D_{d}=\tilde{m}_{d}-\tilde{m}_{s},\quad\quad C_d=1.
\end{eqnarray}
For this texture the quark mixing matrix ${\bf V}$ is a function of three quark mass ratios ${m_u}/{m_t},~{m_c}/{m_t},~{m_d}/{m_s}$ and the CP violation phase $\Phi^*=90^{\circ}$.
%%%%%%%%%%%%%%%%%%%%%%%%%%%%%%%%%%%%%%%%%%%%%%%%%%%%%%%%%%%%%%%%%%%%%%%
As in the first case, the value of ${m_u}$ obtained from ${m_u}/{m_t}$ agrees with the value of $m_u$ at the top scale found in \cite{ref:39} and \cite{ref:1}, the quark mass ratios ${m_d}/{m_s}$ and ${m_c}/{m_t}$ are close to the lower  allowed limit as computed from Eqs. (\ref{eq:5.40}). These values for the quark mass ratios are in very good agreement with the phenomenological relations \cite{ref:26} given in Eq. (\ref{eq:5.58}).\\
In order to have an estimation of the sensitivity of $|V^{th}_{ij}|$ to the uncertainty in the values of the quark mass ratios, we computed the allowed values of $|V^{th}_{ij}|$, corresponding to the range of values of the quark mass ratios given in Table \ref{tab:2}, keeping $\Phi$ fixed at the value $\Phi=90^{\circ}$. The results of the $\chi^2$ fit of the theoretical expressions for $|{V}^{th}_{ij}|$, to the experimentally determined quantities $|{V}^{exp}_{ij}|$ gives the one-standard-deviation ranges shown in Tab. \ref{tab:3}. 
The ``best fit'' values of the theoretical expressions for ${V}^{th}_{ij}$, using the central values of the quark mass ratios from Table \ref{tab:2} are given in Table \ref{tab:4}.\\
As is apparent from Tables \ref{tab:3} and \ref{tab:4}, the agreement between computed and experimental values of all entries in the mixing matrix is very good. In the three cases, the one sigma estimated range of variation of the moduli, the computed values in the four entries of the upper left corner of the matrix  $|{\bf V}^{th}|$ are in agreement with the error bands of the corresponding entries of the matrix of the experimentally determined values of the moduli $|{\bf V}^{exp}|$. The estimated range of variation in the computed values of the entries in the third column and the third row of $|{V}^{th}_{ij}|$ is comparable with the error band of the corresponding entries in the matrix of experimentally determined values of the moduli, with the exception of the element $|V^{th}_{td}|$ in which case the one sigma estimated range of variation due to the uncertainty in the values of the quark mass ratios is significantly smaller than the error band in the experimentally determined value of $|V^{exp}_{td}|$. 

\begin{figure}[tbp]
\begin{center}
\setlength{\unitlength}{0.95cm}
\setlength{\fboxsep}{0cm}
\begin{picture}(14,19)
\put(-1,11.5){$\beta$}
\put(1.5,6.7){${\small m_u/m_t}$}
\put(0,7){\begin{picture}(5,5)\put(-0.9,-0.4){\piccie{6.0cm}{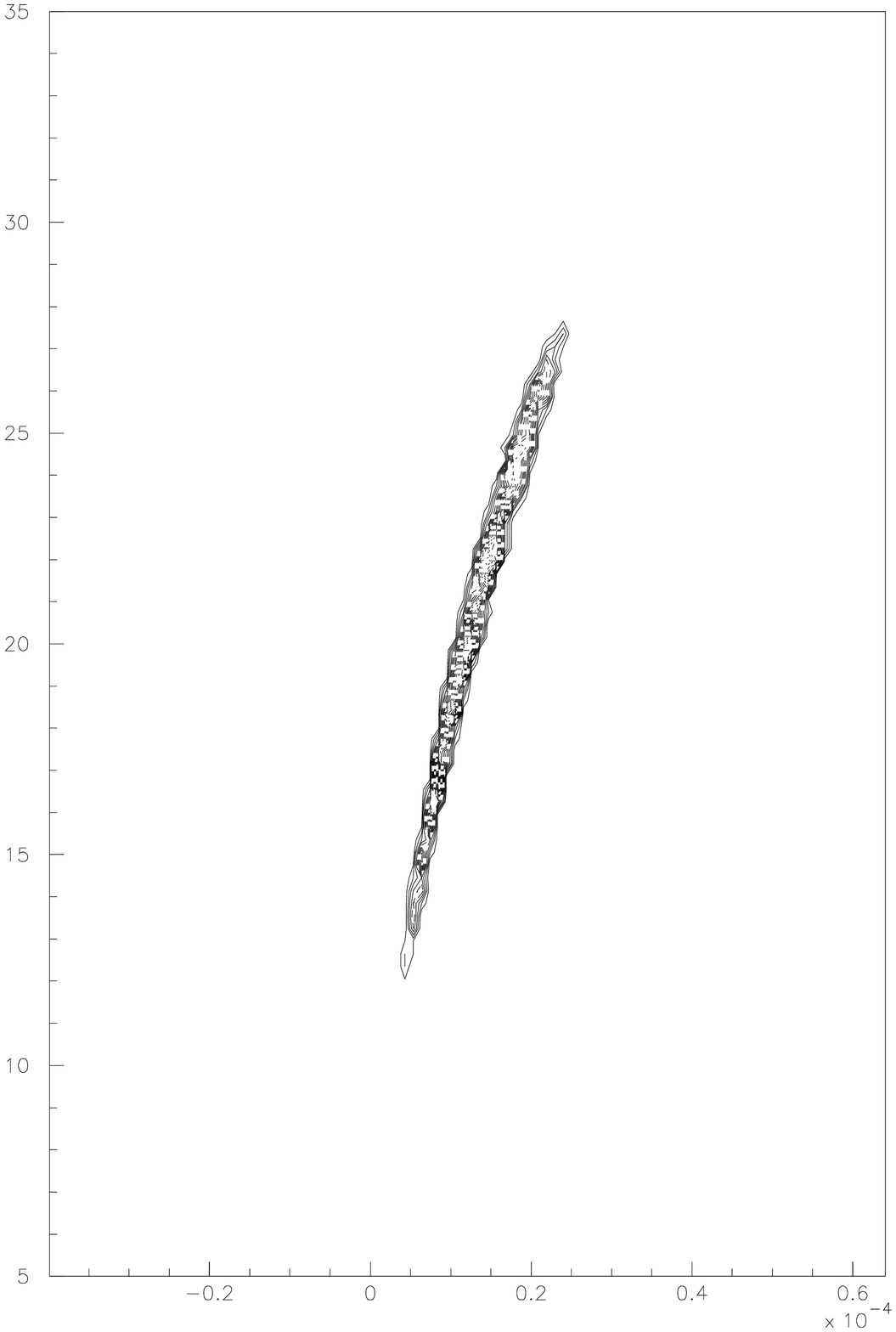}}\end{picture}}
\put(6,11.5){$\beta$}
\put(8.5,6.7){$\alpha$}
\put(7,7){\begin{picture}(5,5)\put(-0.9,-0.4){\piccie{6.0cm}{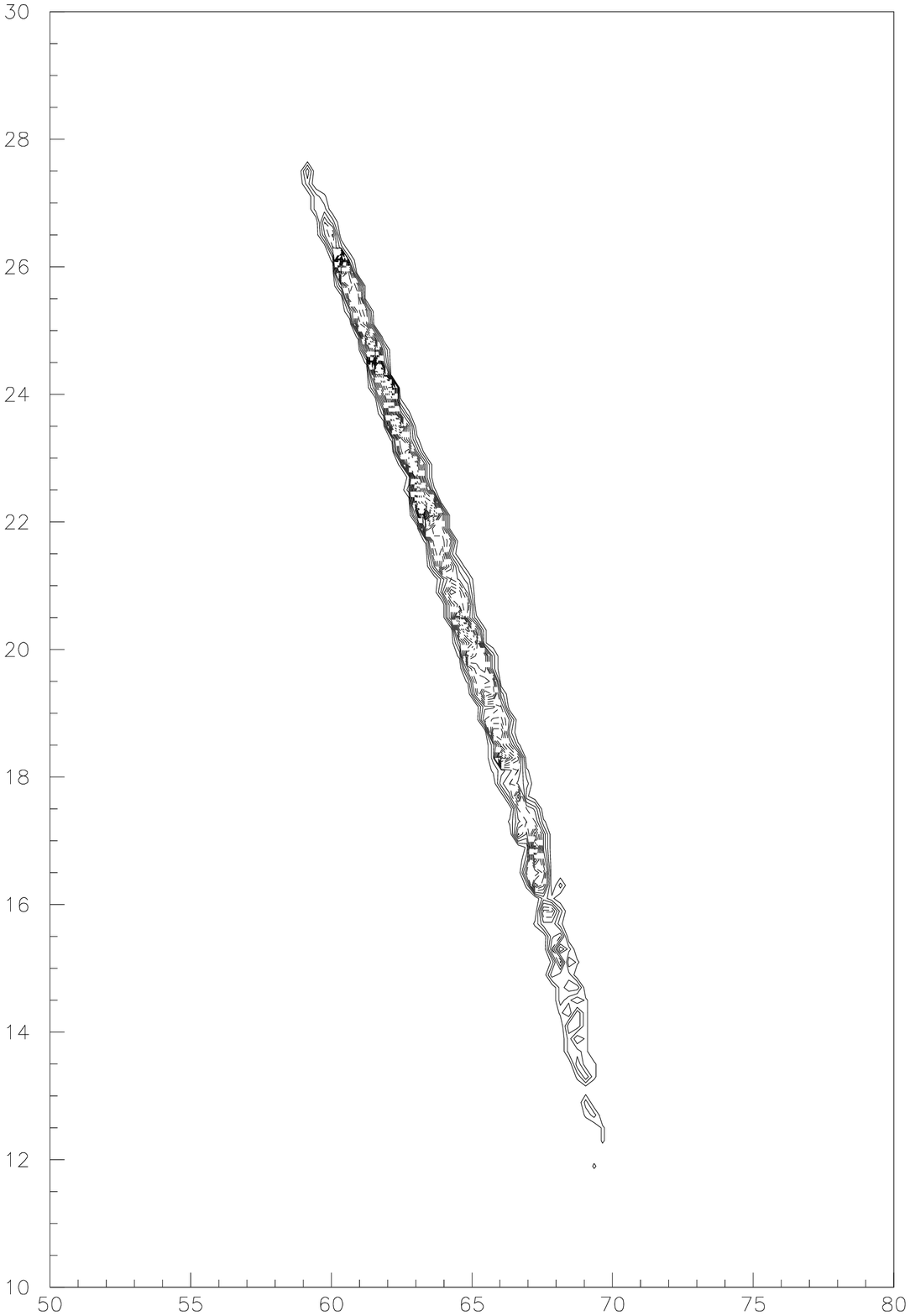}}\end{picture}}
\put(-1,2.5){$\gamma$}
\put(1.5,-2.3){$\alpha$}
\put(0,-2){\begin{picture}(5,5)\put(-0.9,-0.4){\piccie{6.0cm}{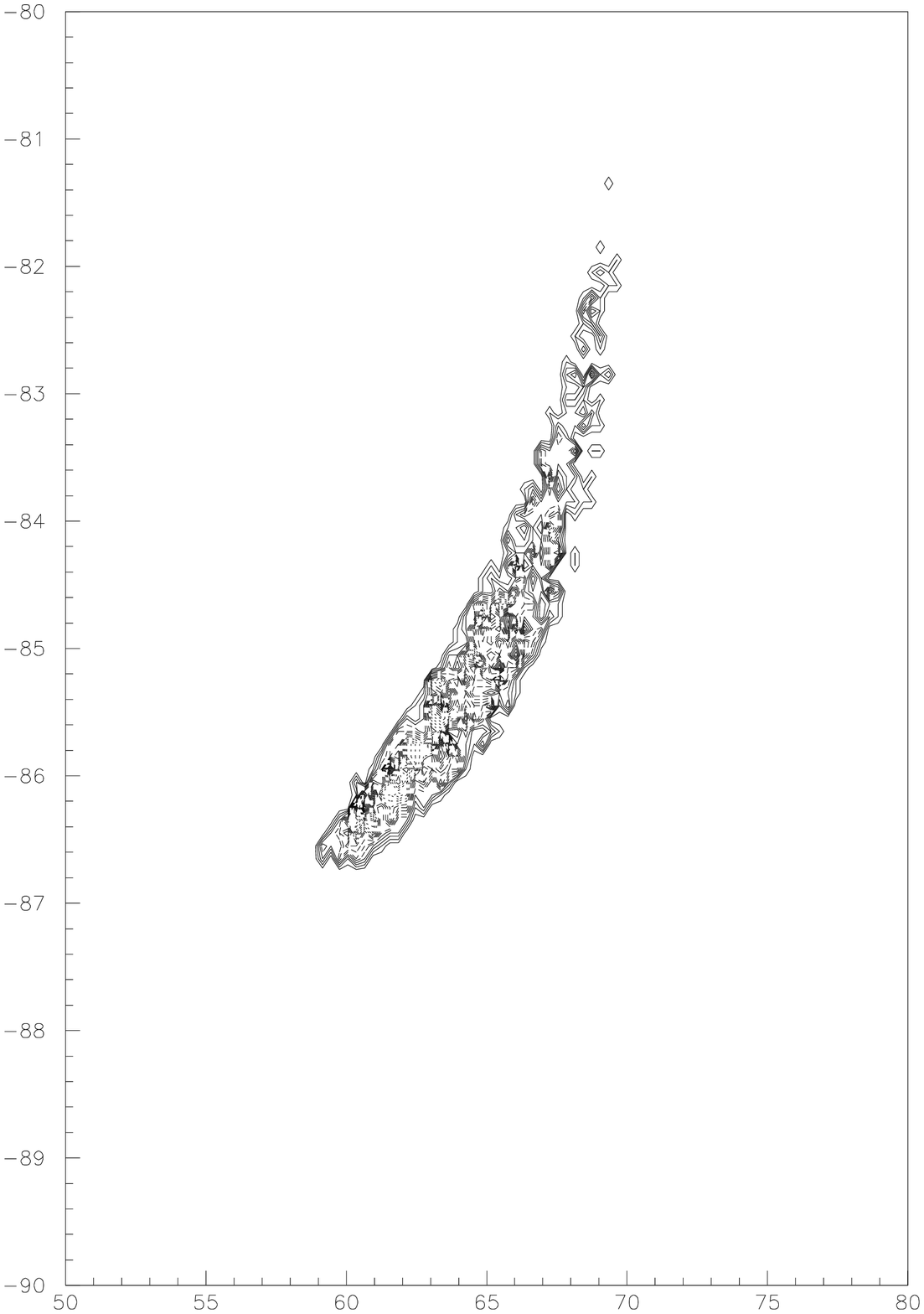}}\end{picture}}
\put(5.6,2.5){${\small \frac{|{\bf V}_{ub}|}{|{\bf V}_{cb}|}}$}
\put(8.5,-2.3){${\small m_u/m_t}$}
\put(7,-2){\begin{picture}(5,5)\put(-0.9,-0.4){\piccie{6.0cm}{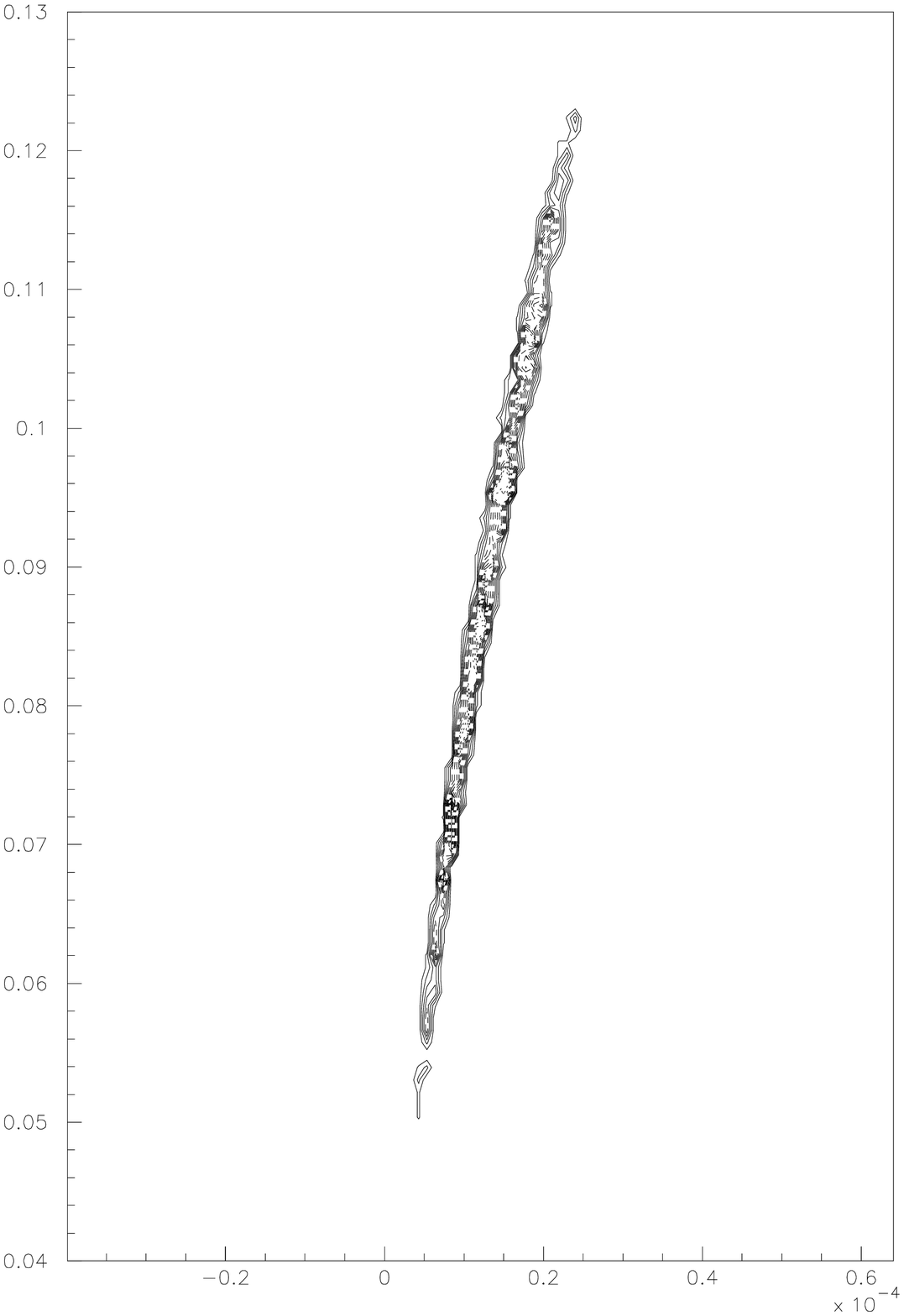}}\end{picture}}
\end{picture}
\end{center}
\vspace{1.cm}\bigskip\bigskip\bigskip
\caption{The one-standard-deviation range of the angles $\beta$, $\gamma$ and the ratio $|V_{ub}|/|V_{cb}|$ computed from the mass pattern III are shown as function of the ratio $mu/mt$. The $\left( \alpha, \beta \right)$ correlation in this mass pattern is also shown.}
\label{fig:2}
\end{figure}

\begin{figure}[tbp]
\begin{center}
\setlength{\unitlength}{0.95cm}
\setlength{\fboxsep}{0cm}
\begin{picture}(14,19)
\put(-1,11.5){$\beta$}
\put(1.5,6.7){${\small m_u/m_t}$}
\put(0,7){\begin{picture}(5,5)\put(-0.9,-0.4){\piccie{6.0cm}{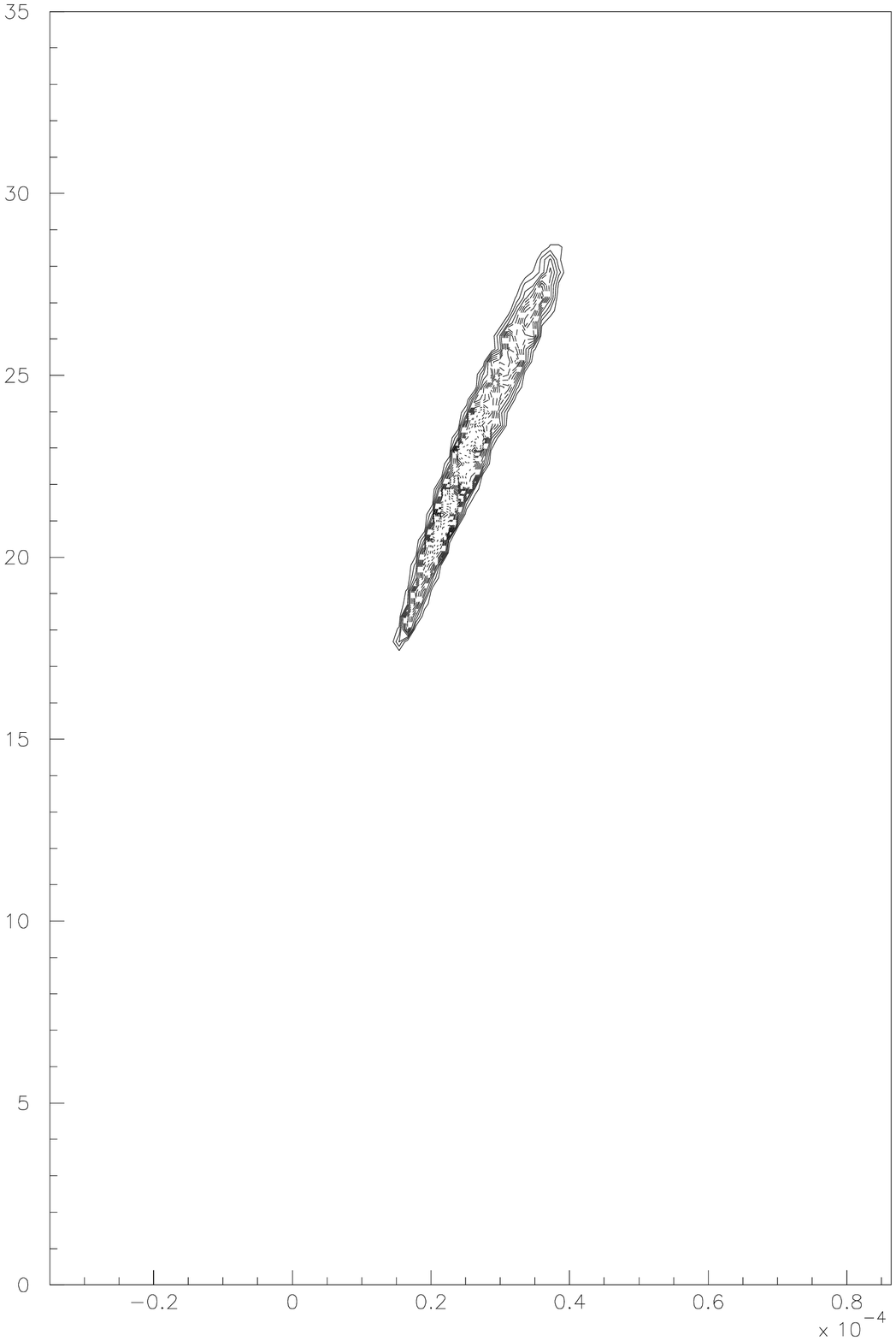}}\end{picture}}
\put(6,11.5){$\beta$}
\put(8.5,6.7){${\alpha}$}
\put(7,7){\begin{picture}(5,5)\put(-0.9,-0.4){\piccie{6.0cm}{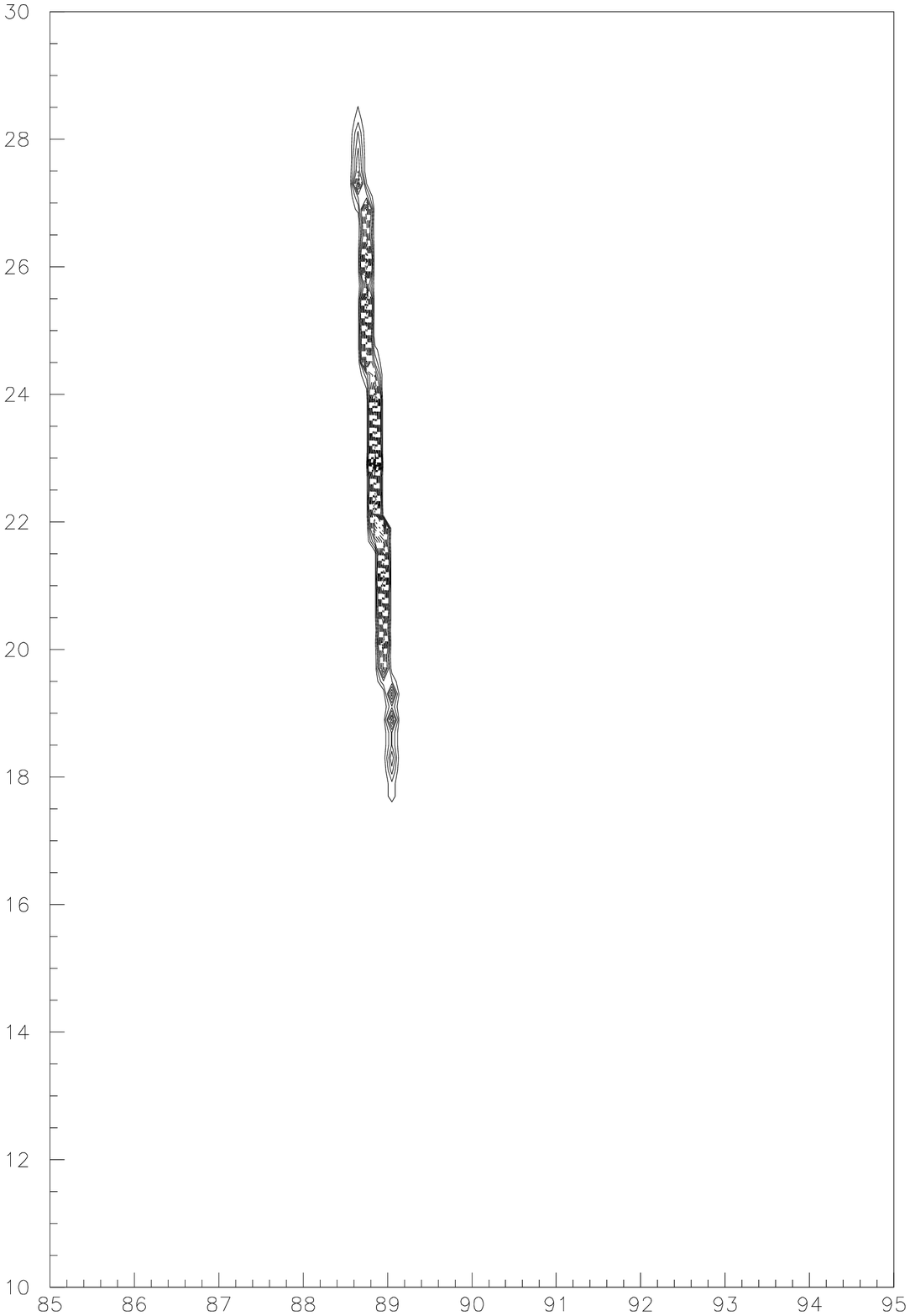}}\end{picture}}
\put(-1,2.5){$\gamma$}
\put(1.5,-2.3){${\alpha}$}
\put(0,-2){\begin{picture}(5,5)\put(-0.9,-0.4){\piccie{6.0cm}{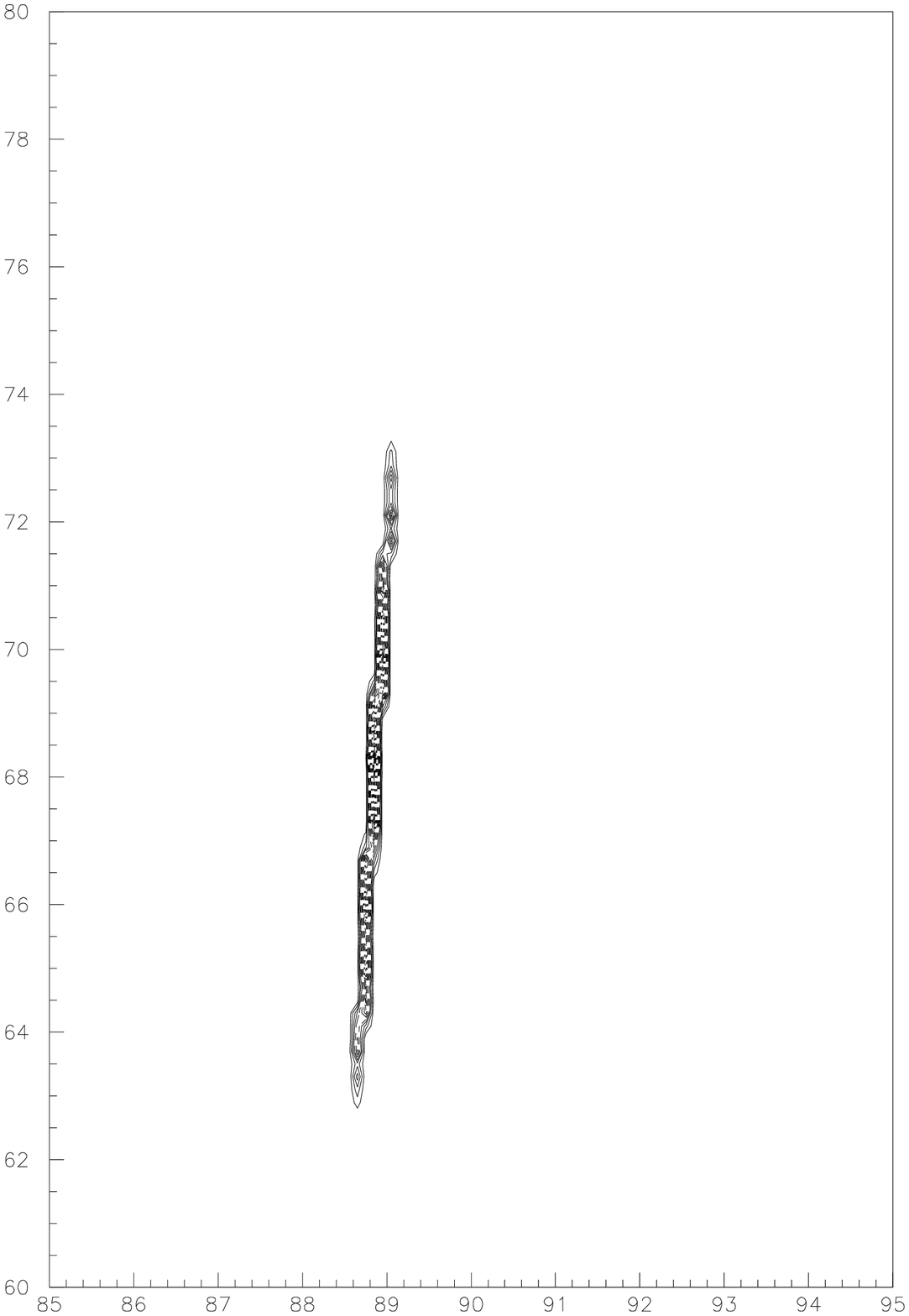}}\end{picture}}
\put(5.6,2.5){${\small \frac{|{\bf V}_{ub}|}{|{\bf V}_{cb}|}}$}
\put(8.5,-2.3){${\small m_u/m_t}$}
\put(7,-2){\begin{picture}(5,5)\put(-0.9,-0.4){\piccie{6.0cm}{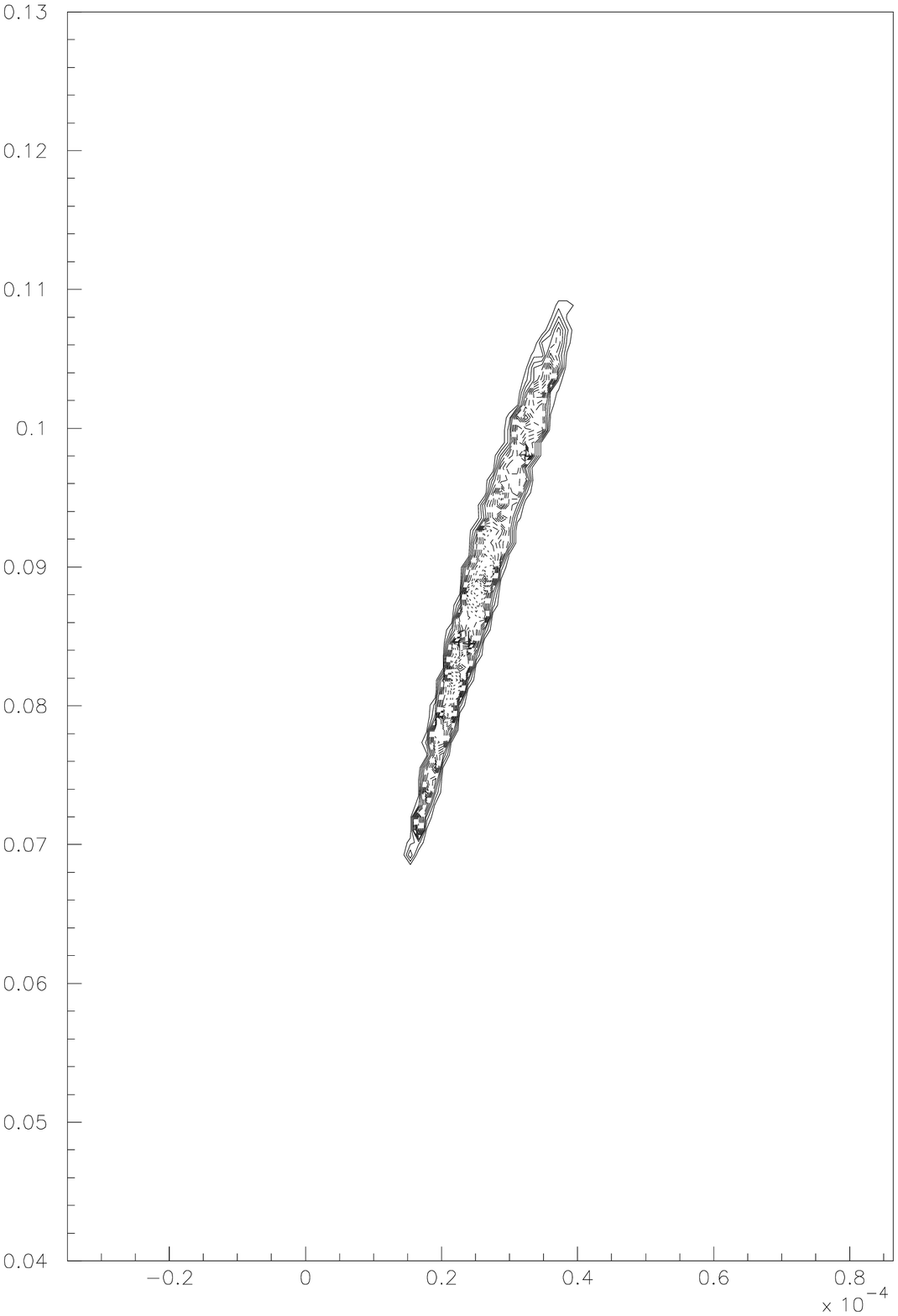}}\end{picture}}
\end{picture}
\end{center}
\vspace{1.cm}\bigskip\bigskip\bigskip
\caption{The one-standard-deviation range of the angles $\beta$, $\gamma$ and the ratio $|V_{ub}|/|V_{cb}|$ computed from the mass pattern IV are shown as function of the ratio $mu/mt$. The $\left( \alpha, \beta \right)$ correlation in this mass pattern is also shown.}
\label{fig:3}
\end{figure}

 The phenomenologically allowed quark mass patterns can be distinguished from the different predicted value for the quark mixing matrix element $|V^{exp}_{td}|$  and hence from the different allowed values of the CP angles $\alpha$, $\beta$ and $\gamma$ of the unitarity triangle. The one-standard-deviation computed range of values of $\alpha$, $\beta$ and $\gamma$, corresponding to the range of values of the mass ratios given in Table \ref{tab:2} with $\Phi$ fixed at the value $\Phi^*=90^{\circ}$ are given in Table \ref{tab:5}.
The uncertainty in the values of the quark mass ratios is reflected in the shape of the unitarity triangle. Thus, different shapes of the unitarity triangle are equivalent to different values of the CP angles. The one-sigma regions of the CP angle $\beta$ are shown in Figs. \ref{fig:1}-\ref{fig:3} as function of the quark mass ratios $m_u/m_t$ and $m_u/m_c$ for the different patterns.  In Figs. \ref{fig:1}-\ref{fig:3} we also illustrate the correlations between the CP angles.

Recent determinations of $\alpha$, $\beta$ and $\gamma$ have been done in \cite{ref:33}-\cite{ref:40}.
It is interesting to compare our results with those of Ali and London \cite{ref:33}, who have studied the profile of the unitarity triangle in the Standard Model and in several variants of the MSSM characterized by a single phase in the quark mixing matrix. According to these authors \cite{ref:33}, an estimation of the range of values of the three inner angles of the 
unitarity triangle, compatible with the experimental information on the 
absolute values of the matrix elements ${\bf V}^{exp}$, in the Standard Model is: $77^{\circ}\leq\alpha\leq 127^{\circ}$, $14^{\circ}\leq\beta\leq 35^{\circ}$, and $34^{\circ}\leq\gamma\leq 81^{\circ}$ at $95\%$ CL.\\

We present in Table \ref{tab:5} the one standard deviation computed ranges of $\alpha$, $\beta$ and $\gamma$ in the three mentioned quark mass textures, together with the corresponding ranges obtained indirectly from the CKM unitarity.
The values of $\beta$ obtained in model I, III, and IV and the values of $\alpha$ given in I and IV are in the one-sigma allowed range given by these authors \cite{ref:33}, while $\alpha$ in model III lies outside the one-sigma range, but is within the  $95\%$ CL range given by  Ali and London \cite{ref:33}. One espects that at the B factories and hadron Colliders the CP-violating angles  $\alpha$, $\beta$ and $\gamma$ will be directly and precisely measured. We see from Table \ref{tab:5} that whereas $\beta$ turns out to be quite similar in the three models listed, precise determination of  $\alpha$ and $\gamma$ may lead to a discrimination among the three models.
%%%%%%%%%%%%%%%%%%%%%%%%%%%%%%%%%%%%%%%%%%%%%%%%%%%%%%%%%%
\section{Conclusions}\label{h3}
The task of searching for phenomenologically viable quark mass textures derived from the condition of maximal Jarlskog invariant $J$ and maximal CP violating phase $\Phi$ is proposed. In order to conduct these searches more effectively, we have suggested in this paper the idea of democratic CP violation from flavor permutational symmetry breaking in the Hermitian mass matrices as an organizing principle.
The conditions of maximal Jarlskog invariant $J$ and  maximal CP violating phase $\Phi^*=90^{\circ}$, encoded in Eq. (\ref{eq:1.39})
 are satisfied by taking particular zero values for the Hermitian quark mass matrix entries.\\
When the CP violating phase takes the values $\Phi^*=90^{\circ}$, all of the elements of the quark mixing matrix moduli $|{\bf V}_{ij}|^2$ have an inflection point and none of them is preferred by Nature, that is, democratic CP violation is realized in Nature. We have shown that the CP violating phase $\Phi$ does not run with energy [Eq. (\ref{eq:5.35})], and the running of the Jarlskog invariant is described by Eqs. (\ref{eq:5.37}) and (\ref{eq:5.39})

From this set of equations, it follows that the zeros imposed at some high energy scale Q for the constrains  $T_1(Q)=0$ or $T_2(Q)=0$ are protected during the renormalization group evolution of the CKM matrix and the masses.

The zeros of the quark mass textures ${\bf M}_u$ and ${\bf M}_d$ in Eqs. (\ref{eq:5.5}-\ref{eq:5.23}) derived from the condition of maximal Jarlskog invariant and maximal CP violating phase $\Phi$ do not give up the parallelism between the structures of ${\bf M}_u$ and ${\bf M}_d$. For this reason, mass paterns with parallel structures of ${\bf M}_u$ and ${\bf M}_d$ could not be a good candidate for an underlying quark mass texture theory as it is supposed in some resent works \cite{ref:1}, \cite{ref:26}. Furthermore, we have argued that maximal Jarlskog invariant $J$ and democratic CP violation with maximal CP violating phase $\Phi$ is the underlying theoretical assumption which leads to the structure of phenomenologically viable RRR \cite{ref:39} quark mass textures.\\
We have evaluated the quark mixing matrix ${\bf V}^{th}$ directly from the hermitian quark mass matrices ${\bf M}_q$. For the quark mass values, taken at the top quark mass scale given in Eq. (\ref{eq:5.40}), only the phenomenologically allowed quark mass patterns I, III and IV in Table \ref{tab:1} give a CKM mixing matrix ${\bf V}^{th}$ in agreement with experiment. As is apparent from Table \ref{tab:3}, these patterns can be distinguished from the predicted value of the quark mixing matrix element $|V_{td}|$. 
We have also performed an analysis of the unitarity triangle. The resulting allowed CP angles of the unitarity triangles in Table \ref{tab:5} obtained from each mass pattern are very constrained. Hence, these patterns can also be distinguished from the different allowed values of the CP angles $\alpha$ or $\gamma$. In the near future, CP-violating asymmetries in B decays will be measured at B-factories and hadron colliders. Such measurements will give us crucial information about the interior angles $\alpha$, $\beta$ and $\gamma$ of the unitarity triangle. If we are lucky, the predictions from one of these patterns will be confirmed pointing to the texture preferred by Nature.

In conclusion, the hypothesis of maximal Jarlskog invariant and democratic CP violation fixes the CP violating phase $\Phi$ to $\Phi=\Phi^*=90^{\circ}$ and the fact that the phenomenologically viable quark mass textures for the quarks can be obtained from  the conditions of maximal Jarlskog invariant $J$ and maximal CP violating phase $\Phi^*$ gives evidence in favor of this hypothesis.

All the analysis for the maximal Jarlskog invariant and democratic maximal CP violating phase $\Phi$ apply also to the lepton sector and the allowed mass patterns for the lepton sector can simply be obtained by using the Georgi-Jarlskog Ansatz \cite{ref:43} taking all the elements of $Y-{\it l}$ as the same as for $Y-{\it d}$ except the $(2,~2)$ element which is multiplied by the factor 3. It is clear from this assumption, that once we know the right quark mass pattern, we can make predictions for the neutrino mass ratios and lepton mixing matrix.

%%%%%%%%%%%%%%%%%%%%%%%%%%%%%%%%%%%%%%%%%%%%%%%%%%%%%%%%%%%%%%%%%%5
%%%%%%%%%%%%%%%%%%%%%%%%%%%%%%%%%%.%%%%%%%%%%%%%%%%%%%%%%%%%%%%%%%%%%%%%%
\section*{Acknowledgments}
I would like to specially thank Prof. Ahmed Ali for supervising this work and reading the manuscript. I also thank Nabil Ghodbane for his invaluable help with MINUIT and Dimitri Ozerov for his invaluable help with PAW and reading the manuscript. This work was partly supported by an ICSC-World Laboratory Bjorn Wiik Scholarship and by CONACYT (M\'exico) under contract 000441.

\begin{table}
\begin{center}
\begin{tabular}{|c|c|c|c|c|c|}
\hline
RRR patterns \cite{ref:39}  &  I &  II & III & IV & V \\
\hline
${\bf M}_u$ & $\pmatrix{
0 & A & 0 \cr
A & D & 0 \cr
0 & 0 & C \cr
}$  & $\pmatrix{
0 & A & 0 \cr
A & 0 & B \cr
0 & B & C \cr
}$ &$\pmatrix{
0 & 0 & F \cr
0 & D & 0 \cr
F & 0 & C \cr
}$  
& $\pmatrix{
0 & A & 0 \cr
A & D & B \cr
0 & B & C \cr
}$ & $\pmatrix{
0 & 0 & F \cr
0 & D & B \cr
F & B & C \cr
}$\\
\hline
${\bf M}_d$ & $\pmatrix{
0 & A & 0 \cr
A & D & B \cr
0 & B & C \cr
}$  & $\pmatrix{
0 & A & 0 \cr
A & D & B \cr
0 & B & C \cr
}$ & $\pmatrix{
0 & A & 0 \cr
A & D & B \cr
0 & B & C \cr
}$ & $\pmatrix{
0 & A & 0 \cr
A & D & 0 \cr
0 & 0 & C \cr
}$  & $\pmatrix{
0 & A & 0 \cr
A & D & 0 \cr
0 & 0 & C \cr
}$ \\
\hline
\end{tabular}
\caption{Phenomenologically allowed patterns of the Hermitian quark mass matrices in the Ramond-Robert-Ross approach.} 
\label{tab:1}
\end{center}
\end{table}

\begin{table}
\begin{center}
\begin{tabular}{|c|c|c|c|c|}
\hline
  & I &  III & IV & Fusaoka-Koide \cite{ref:25} \\
\hline
${m_u}/{m_t}$ &   & $(1.2\pm 1.0)\times 10^{-5}$ &  $(2.6\pm 1.2)\times 10^{-5}$ &  $(1.30\pm 0.28)\times 10^{-5}$   \\
\hline
${m_c}/{m_t}$ &  &  &  $(3.28\pm 0.33)\times 10^{-3}$ &  $(3.78\pm 0.39)\times 10^{-3}$  \\
\hline
${m_u}/{m_c}$ & $(7.43\pm 6.45)\times 10^{-3}$ & &  &  $(3.4\pm 1.13)\times 10^{-3} $ \\
\hline
${m_d}/{m_b}$ & $(1.695\pm 0.174)\times 10^{-3}$  & $(1.68\pm 0.17)\times 10^{-3}$ &  &  $(1.58\pm 0.30)\times 10^{-3}$   \\
\hline
${m_s}/{m_b}$ & $(3.8\pm 0.67)\times 10^{-2}$ & $(3.22\pm 0.34)\times 10^{-2}$ &  &  $(3.14\pm 0.58)\times 10^{-2}$  \\
\hline
${m_d}/{m_s}$ &  & &  $(4.42\pm 0.375)\times 10^{-2}$ &  $(5.02\pm 1.65)\times 10^{-2}$ \\
\hline
\end{tabular}
\caption{One standard deviation ranges of the running quark mass ratios in the three RRR quark mass matrices listed in Table \ref{tab:1}.}
\label{tab:2}
\end{center}
\end{table}

\begin{table}
\begin{center}
\begin{tabular}{|c|c|c|c|c|}
\hline
  &  I &  III & IV & PDG \cite{ref:3}\\
\hline
$|V_{ud}|$ & 0.9745-0.9754  & 0.9746-0.9753 & 0.9746-0.9754 & 0.9742-0.9757 \\
\hline
$|V_{us}|$ & 0.2205-0.224  & 0.2207-0.224 &  0.2205-0.224 & 0.219-0.226 \\
\hline
$|V_{ub}|$ & 0.002-0.005  & 0.002-0.005 & 0.0028-0.0044 & 0.002-0.005   \\
\hline
$|V_{cd}|$ & 0.2205-0.2238 & 0.2205-0.2238 & 0.2205-0.2238 & 0.219-0.225 \\
\hline
$|V_{cs}|$ & 0.9738-0.9745   & 0.9738-0.9745 & 0.9738-0.9745 & 0.9734-0.9749 \\
\hline
$|V_{cb}|$ & 0.038-0.042  & 0.0382-0.0422 &  0.0382-0.0422 & 0.037-0.043  \\
\hline
$|V_{td}|$ & 0.0078-0.0096  & 0.0094-0.0103  &  0.0076-0.009 & 0.004-0.014   \\
\hline
$|V_{ts}|$ & 0.0375-0.0415  & 0.0375-0.041 &  0.0375-0.0415 & 0.035-0.043  \\
\hline
$|V_{tb}|$ & 0.9991-0.99926  & 0.9991-0.99926 &  0.9991-0.99926 & 0.9991-0.9993  \\
\hline
$\frac{|V_{ub}|}{|V_{cb}|}$ & $0.087\pm 0.034$ & $0.087\pm 0.034$ & $0.089\pm 0.021$ & $0.09\pm 0.025$  \\
\hline
$|J|$ & $(2.9\pm 1.1)\times 10^{-5}$ & $(3.0\pm 1.2)\times 10^{-5}$ & $(2.9\pm 0.7)\times 10^{-5}$ &   \\
\hline
\end{tabular}
\caption{One-standard-deviation ranges of $|V^{th}_{ij}|$ in the three RRR quark mass matrices listed in Table \ref{tab:1} compared with the PDG values.}
    \label{tab:3}
\end{center}
\end{table}

\begin{table}
\begin{center}
\begin{tabular}{|c|c|c|c|c|}
\hline
  &  I &  III & IV & $|V^{exp}_{ij}|$ \\
\hline
$V_{ud}$ & $0.975 e^{-i}$ & 0.975 & $0.975e^{-i}$ & 0.97495 \\
\hline
$V_{us}$ & $0.222e^{i22^{\circ}}$ & $0.222e^{i180^{\circ}}$ &  $0.2224e^{i23^{\circ}}$ & 0.2225 \\
\hline
$V_{ub}$ & $0.0035e^{-i85^{\circ}}$  & $0.0035e^{-i85^{\circ}}$ & $0.0036e^{-i90^{\circ}}$ & 0.0035  \\
\hline
$V_{cd}$ & $0.222e^{i68^{\circ}}$ & $0.222e^{-i90^{\circ}}$& $0.2222e^{i67^{\circ}}$ & 0.222 \\
\hline
$V_{cs}$ & $0.974e^{-i89^{\circ}}$  & $0.974e^{-i90^{\circ}}$ & $0.974e^{-i89^{\circ}}$ & 0.97415 \\
\hline
$V_{cb}$ & $0.040e^{i90^{\circ}}$ & $0.040e^{-i90^{\circ}}$ &  $0.04e^{-i90^{\circ}}$ & 0.040  \\
\hline
$V_{td}$ & $0.0087e^{-i90^{\circ}}$  & $0.0099 e^{-i70^{\circ}}$ & $0.0083e^{-i90^{\circ}}$ & 0.009   \\
\hline
$V_{ts}$ & $0.0395e^{-i90^{\circ}}$  & $0.039e^{i89^{\circ}}$ &  $0.0396e^{i90^{\circ}}$ & 0.039  \\
\hline
$V_{tb}$ & $0.9992e^{-i90^{\circ}}$  & $0.9992e^{i90^{\circ}}$ &  $0.9992e^{-i90^{\circ}}$ & 0.99915 \\
\hline
\end{tabular}
\end{center}
\caption{Central values of $V^{th}_{ij}$ in models I, III and IV (from Table \ref{tab:1}) and the central values of $|V_{ij}|$ from experiments. }
\label{tab:4}
\end{table}

\begin{table}
\begin{center}
\begin{tabular}{|c|c|c|c|c|}
\hline
Solution  & I &  III & IV & Ali-London \cite{ref:33} \\
\hline
$\alpha$ & $(84\pm 3)^{\circ}$  & $(65\pm 7)^{\circ}$ &  $(89\pm 0.4)^{\circ}$ &  $86^{\circ} \le \alpha \le 111^{\circ}$   \\
\hline
$\beta$ & $(22\pm 10)^{\circ}$ & $(20\pm 8)^{\circ}$ &  $(23\pm 7)^{\circ}$ &  $18^{\circ} \le \beta \le 28^{\circ}$  \\
\hline
$\gamma$ & $(74\pm 12)^{\circ}$ & $(95\pm 3)^{\circ}$ &  $(68\pm 6)^{\circ}$ &  $49^{\circ} \le \gamma \le 72^{\circ}$ \\
\hline
\end{tabular}
\end{center}
\caption{One-standard-deviation ranges of $\alpha$, $\beta$ and $\gamma$  in the texture models I, III and IV in Table \ref{tab:1} and their comparison from the corresponding indirect determinations using the CKM unitarity.}
\label{tab:5}
\end{table}

\end{document}